\documentclass[letterpaper,11pt]{article}

\newif\ifdraft
\ifdefined\draftmode
  \drafttrue
\else
  \draftfalse
\fi

\usepackage[english]{babel}
\usepackage[utf8]{inputenc}
\usepackage[T1]{fontenc}
\usepackage{lmodern}
\usepackage{microtype}
\usepackage{amsmath,amssymb,amsthm,mathtools}
\usepackage{mathrsfs}
\usepackage{enumitem}
\usepackage{graphicx}
\usepackage{array}
\usepackage{authblk}
\usepackage[table]{xcolor}
\usepackage{float}
\usepackage{booktabs,multirow}
\usepackage{nicefrac}
\usepackage{xspace}
\usepackage[backref=page]{hyperref}
\usepackage{natbib}

\usepackage[letterpaper,margin=1in]{geometry}
\hypersetup{hidelinks}
\setlist{nosep,leftmargin=*}
\newcommand{\aptas}{\textrm{APTAS}}

\newtheorem*{theorem*}{Theorem}
\newtheorem*{corollary*}{Corollary}

\usepackage{pgfplots}
\usepackage{tikz}
\usepackage[caption=false]{subfig}
\usetikzlibrary{arrows.meta,patterns,patterns.meta,decorations.pathreplacing,
  positioning,decorations.text,decorations.pathmorphing,calc}
\pgfplotsset{compat=newest}

\floatstyle{ruled}
\newfloat{algorithm}{htbp}{loa}
\floatname{algorithm}{Algorithm}
\newcounter{algorithmline}

\definecolor{aptasblue}{RGB}{31,92,153}
\definecolor{aptasbluelight}{RGB}{190,218,242}
\definecolor{aptasteal}{RGB}{0,112,105}
\definecolor{aptasteallight}{RGB}{180,222,212}
\definecolor{aptasamber}{RGB}{184,112,0}
\definecolor{aptasamberlight}{RGB}{246,211,136}
\definecolor{aptasred}{RGB}{172,48,48}
\definecolor{aptasredlight}{RGB}{239,184,180}
\definecolor{aptasgray}{gray}{0.42}
\definecolor{aptasgraylight}{gray}{0.94}

\tikzset{
  figframe/.style={draw=aptasgray,line width=0.8pt,rounded corners=1.5pt},
  figpanel/.style={figframe,fill=aptasbluelight!24},
  figgrid/.style={draw=aptasblue!72!black,line width=0.62pt},
  figfine/.style={draw=aptasblue!58!black,densely dotted,line width=0.48pt},
  figarrow/.style={-{Latex[length=2.0mm,width=1.35mm]},line width=0.85pt,draw=aptasblue},
  figmap/.style={figarrow,line width=1.0pt},
  figtitle/.style={font=\small\bfseries,align=center,text=black!88},
  figlabel/.style={font=\footnotesize,align=center,text=black!82},
  fignote/.style={font=\scriptsize,align=center,text=black!72},
  currentpoly/.style={draw=aptasteal!80!black,fill=aptasteallight,line width=1.05pt},
  reservedcell/.style={draw=aptasblue!80!black,fill=aptasbluelight,pattern=north west lines,pattern color=black!58,line width=0.9pt},
  activecell/.style={draw=aptasblue,double=white,double distance=0.35pt,line width=0.85pt},
  inactivecell/.style={draw=black!62,dashed,fill=aptasgraylight,line width=0.75pt},
  discardcell/.style={fill=aptasredlight,pattern=north east lines,pattern color=black!58,line width=0.85pt},
  problemcell/.style={draw=aptasred!80!black,fill=aptasredlight,pattern=north east lines,pattern color=black!62,line width=1.0pt},
  retainedband/.style={draw=aptasblue!80!black,fill=white,line width=0.9pt,rounded corners=2.5pt},
  rootband/.style={draw=aptasamber!80!black,dashed,rounded corners=2.5pt,fill=white,line width=0.7pt},
  separatorband/.style={draw=aptasred!80!black,fill=aptasredlight,pattern=north east lines,pattern color=aptasredlight,rounded corners=2.5pt,line width=0.85pt},
  chainbox/.style={rounded corners=3pt,line width=0.8pt},
  chainboxplain/.style={chainbox,draw=aptasblue!70!black,fill=aptasbluelight!50},
  chainboxmarked/.style={chainbox,draw=aptasamber!80!black,fill=aptasamberlight!40},
  pics/nonconvexPolygon/.style={code={
      \draw[pic actions]
      (-0.22,-0.10)--(-0.13,0.20)--(0.17,0.17)--(0.03,0.02)--(0.23,-0.10)--(-0.01,-0.17)--cycle;
    }}
}

\newtheorem{theorem}{Theorem}
\newtheorem{lemma}{Lemma}
\newtheorem{proposition}{Proposition}
\newtheorem{corollary}{Corollary}
\newtheorem{definition}{Definition}

\newcommand{\OPT}{\mathrm{OPT}}
\newcommand{\area}{\operatorname{Area}}

\newcommand{\eps}{\varepsilon}
\newcommand{\calP}{\mathcal{P}}
\newcommand{\callB}{\mathcal{B}}
\newcommand{\gridrate}{\Pi}
\newcommand{\Kloc}{M}
\newcommand{\Rrep}{K_{\mathrm{rep}}}
\newcommand{\Bchild}{B_{\mathrm{child}}}
\newcommand{\Hpat}{H_{\mathrm{cfg}}}
\newcommand{\Mgrid}{M_{\mathrm{grid}}}
\newcommand{\Sclass}{\mathscr{C}}
\newcommand{\Ncls}{K}

\newcommand{\Zcut}{\mathcal{Z}_{\mathrm{cut}}}
\newcommand{\Zfringe}{\mathcal{Z}_{\mathrm{fringe}}}
\newcommand{\Zprob}{\mathcal{Z}_{\mathrm{prob}}}
\newcommand{\Zexc}{\mathcal{Z}_{\mathrm{exc}}}

\newlength{\RoundedBoxWidth}
\newsavebox{\RoundedBoxContents}

\newenvironment{RoundedBox}[1]{%
  \setlength{\RoundedBoxWidth}{.93\textwidth}%
  \def\boxheading{#1}%
  \begin{lrbox}{\RoundedBoxContents}%
    \begin{minipage}{\RoundedBoxWidth}%
}{%
    \end{minipage}%
  \end{lrbox}%
  \begin{center}%
    \begin{tikzpicture}%
      \node (Text) [draw=black!20,fill=white,rounded corners,inner sep=2ex,text width=\RoundedBoxWidth] {\usebox{\RoundedBoxContents}};%
      \coordinate (HeadingAnchor) at (current bounding box.north west);%
      \node [draw=white,rectangle,inner sep=3pt,anchor=north west,fill=white] at ($(HeadingAnchor)+(6pt,.75em)$) {\boxheading};%
    \end{tikzpicture}%
  \end{center}%
}

\newenvironment{problemstatement}[2][]{%
  \noindent\ignorespaces%
  \parindent=0pt%
  \vspace*{-1.5em}%
  \if\relax\detokenize{#1}\relax%
    \begin{RoundedBox}{\textsc{#2}}%
  \else%
    \begin{RoundedBox}{\textsc{#2} parameterized by~#1}%
  \fi%
  \begin{tabular}{@{\hspace{.1em}} >{\raggedright\arraybackslash}p{2cm} p{\dimexpr\linewidth-2cm-2\tabcolsep-.1em\relax}@{}}%
}{%
  \end{tabular}%
  \end{RoundedBox}%
  \ignorespacesafterend%
}

\ifdraft

\usepackage{todonotes}
\fi

\begin{document}

\title{
  Geometric Bin Packing Beyond Convexity:\\
  Approximation Schemes without Augmentation
}

\author{Elisa Dell'Arriva}
\author{Lucas de Oliveira Silva\thanks{Corresponding author.}}
\author{Fl\'avio Keidi Miyazawa}
\setlength{\affilsep}{1cm}
\affil{
  Institute of Computing, University of Campinas, São Paulo, Brazil
  \authorcr \texttt{\small \{elisa.arriva,lucas.oliveira.silva,fkm\}@ic.unicamp.br}\medskip
}

\date{}
\hypersetup{pageanchor=false}
\begin{titlepage}
\maketitle

\begin{abstract}
    We study the geometric bin packing problem for simple polygonal regions, i.e., non-degenerate, connected, rational regions, in the minimum number of unit-square bins, allowing only translations.
    We even consider packing general non-convex objects, possibly with holes, where we can then pack smaller objects.
    Previous schemes cover squares (MOR'06), circles and other $L_p$-balls (Algorithmica'16), sufficiently skewed rectangles (Algorithmica'23), and convex fat objects with a constant number of distinct shapes under resource augmentation (arXiv'24, extending WAOA'23).
    The result for circles and other $L_p$-balls needs bin augmentation to obtain rational packing realizations.
    For restricted classes of $L$-shaped polygons, Gerlach, Hennies, and Kleist (SoCG'26) give constant worst-case guarantees, while for general \mbox{non-convex} objects, bounding-box reductions give just approximation factors that depend on fatness.
    
    Against this background, we present a general framework that provides asymptotic polynomial-time approximation schemes (APTAS) for bin packing of any fixed class of simple polygonal regions satisfying just two assumptions.
    First, covering any object by the grid cells intersecting its interior in any sufficiently fine square grid, relative to the object's bounding box, increases its area by at most a small fraction.
    Second, the class can be partitioned into a constant number of sequences such that each object in a sequence can be translated to fit inside every preceding object.
    In particular, certain classes of star-shaped polygons and more general polygonal regions, including some with simple holes, satisfy both assumptions.
    In addition, our framework does not require a bound on the objects' boundary complexity.
    We emphasize that our results require no bin augmentation whatsoever and yield explicit, exact rational packings of polynomial encoding length for all input objects.
    
    Finally, we remark that our grid assumption is arguably mild.
    Indeed, for convex objects, we establish equivalence with a fatness bound, whereas for non-convex objects, these assumptions are incomparable.
    Nevertheless, even for rectangles, some hypothesis beyond fatness, or the grid covering assumption, is unavoidable: assuming $\mathrm{P} \ne \mathrm{NP}$, Chlebík and Chlebíková (JDA'09) rule out an APTAS for a family of axis-aligned fat rectangles of uniformly positive area.
\end{abstract}

\thispagestyle{empty}
\end{titlepage}
\hypersetup{pageanchor=true}

\section{Introduction}

In the geometric bin packing problem, we are given a multiset of geometric objects, and the goal is to compute a non-overlapping placement of all objects into the minimum number of identical geometric bins.
The literature on exact formulations and computational heuristics is broad for general objects.
Approximation results, by contrast, are more scarce.
For overviews of this literature, see the surveys of Bennell and Oliveira and Alt~\cite{BO09,Alt16}.
In Part I of a two-part study, Daniels and Milenkovic~\cite{DM97} present an approximate algorithm for translational polygon containment that allows overlaps within a prescribed geometric tolerance, and compare it experimentally with exact algorithms.
In Part II, Milenkovic~\cite{Mil97} gives exact algorithms that decide whether $k$ non-convex polygons fit in a polygonal container by translation and find a placement when one exists, in time polynomial in the total boundary complexity for fixed $k$.
Moreover, related complexity results address exact packing feasibility rather than approximations~\cite{AMS24}.

By contrast, the literature on approximation algorithms has mainly focused on convex objects.
Bansal, Correa, Kenyon, and Sviridenko~\cite{BCKS06} showed that hypercubes admit an APTAS for any fixed dimension $d$.
This, in particular, implies an APTAS for squares.
The same authors showed that for general \mbox{axis-aligned} rectangles, the bin packing problem admits no APTAS unless $\mathrm{P}=\mathrm{NP}$.
Bansal, Caprara, and Sviridenko~\cite{BCS09} gave a polynomial-time approximation for rectangles with an asymptotic ratio of about $1.525$.
Later, Bansal and Khan~\cite{BK14} improved this asymptotic ratio to about $1.405$ in polynomial time for arbitrary rectangles, with or without $90$-degree rotations.
With resource augmentation, arbitrary rectangles can be packed into at most $\OPT$ square bins after increasing each bin side length from $1$ to $1+\gamma$ for every fixed positive $\gamma$~\cite{BCKS06}.
On the other hand, in $2009$, Chleb\'{\i}k and Chleb\'{\i}kov\'a~\cite{CC09} showed that, unless $\mathrm{P}=\mathrm{NP}$, no APTAS exists even for axis-aligned fat rectangles of uniformly positive area.
More recently, Khan and Sharma~\cite{KSh23} gave an APTAS for skewed rectangles, where every rectangle has at least one side sufficiently small as a function of the desired accuracy $\eps$.
For convex polygons, Kurpisz and Suter~\cite{KS23} gave a $5.09$-approximation for the case where the diameter of each polygon is at most $\nicefrac{1}{10}$, and an $O(\nicefrac{1}{\delta})$-approximation when every polygon has width or height at most $1-\delta$.

For circle bin packing, Miyazawa, Pedrosa, Schouery, Sviridenko and Wakabayashi~\cite{MPSSW14} gave an APTAS.
Their result also extends to regular convex polygons and $L_p$-spheres.
For circles, their algorithm obtains rational realizations in bins of width $1$ and height $1+\gamma$, for arbitrarily small $\gamma>0$.
This augmentation, in turn, permits rational approximation of potentially required irrational center coordinates.
They left open whether every instance with rational radii admits a packing using just rational center coordinates in the same bin count.
More broadly, in the full version of~\cite{CDM23}, Chagas, Dell'Arriva and Miyazawa develop a common framework for several geometric packing problems for convex fat objects with a constant number of distinct shapes, including the multiple knapsack problem, the bin packing problem, the strip packing problem, and \mbox{minimum-container} variants.
For bin packing, their result yields an efficient \mbox{resource augmentation} scheme that packs into augmented bins using no more bins than an optimal unaugmented packing and supports features such as multiplicities and rotations.

For non-convex polygons with no diameter restriction, few results provide constant-factor guarantees independent of the objects' aspect ratios.
Only recently, Gerlach, Hennies, and Kleist obtained such bounds for restricted classes of L-shaped polygons~\cite{GHK26}.
Namely, their offline results imply an asymptotic $10$-approximation for symmetric L-shapes of arbitrary size.

\subsection{Our contribution}

In this work, we study the geometric bin packing problem for general connected polygonal regions, possibly non-convex and with holes, using unit-square bins.
We consider the offline, translation-only setting and assume the polygonal regions have rational vertices.
Our main contribution is a framework that yields an APTAS for bin packing of any fixed object class satisfying two assumptions, as stated in Theorem~\ref{thm:abstract-main}.
The assumptions concern approximation by square-grid cells and representative rounding under translation containment.

The first assumption requires that, for any sufficiently fine square grid, cells whose interiors intersect an object's interior cover it, adding at most a prescribed relative excess area.
Concretely, for a fixed object class and any rational relative error $0<\mu<1$, the grid assumption supplies a cell-size bound relative to each object's bounding-square side length, uniform over the class and grid phase, so that the grid-cell cover has area at most $1+\mu$ times the object's area.

The second assumption requires a rounding procedure that, after selecting a few input objects for separate packing, replaces the remaining objects with copies of chosen maxima of \emph{containment groups}, meaning members that contain a translate of every object in the same group.
For fixed class and accuracy, this rounding is computed efficiently, yielding a bounded number of groups at each diameter scale.
Moreover, replacing retained objects with copies of their group maximum must produce copies that can be packed in no more bins than the input.
By choosing the rounding parameters appropriately, we make the bin count of the separate packing an arbitrarily small fraction of the input optimum, up to an instance-independent additive constant.
Definition~\ref{def:roundable} gives the precise guarantees.
Under the grid assumption, a sufficient condition for this rounding is bounded \emph{translation-containment width}: every input can be partitioned into a class-dependent constant number of chains, with every earlier object in a chain containing a translate of every later one. \medskip

We emphasize that our algorithm computes an exact rational realization of the packing, with all objects placed at rational coordinates and without resource augmentation.
Additionally, the number of boundary vertices may even be part of the input.

In particular, we obtain an \aptas{} for rational homothetic copies of any fixed finite collection of rational star-shaped polygons, as stated in Corollary~\ref{cor:homothetic-star-ptas}.
Formally, a \emph{(rational) homothetic copy} of an object is obtained by a positive (rational) scaling followed by a translation, with neither rotation nor reflection.
A polygon is \emph{star-shaped} if it contains a point whose line segment to every point of the polygon remains inside the polygon.
For example, triangles, rectangles, and all convex polygons are star-shaped.
However, non-convex polygons may also be star-shaped.
For example, for homothetic copies of a fixed finite set of symmetric L-shaped objects, we provide an APTAS.

For convex classes, our grid assumption follows from a uniform bound on the ratio of circumradius to inradius, the usual notion of uniform fatness.
Under the additional hypothesis of bounded translation-containment width, we obtain an \aptas{} for any fixed uniformly fat class of rational convex polygons (Corollary~\ref{cor:concrete}).
However, the grid assumption alone is insufficient: Chlebík and Chlebíková~\cite{CC09} showed that, unless $\mathrm{P}=\mathrm{NP}$, no APTAS exists even for fat axis-aligned rectangles.
By contrast, for non-convex regions, fatness and the grid condition are incomparable, as shown in Proposition~\ref{prop:nonconvex-grid-fat}.
More generally, our framework applies to any fixed class of non-convex polygonal regions satisfying the two assumptions, including regions with holes.

\subsection{Techniques}

Fix an accuracy parameter $0<\eps<1$.
We prove that imposing a simplified arrangement on a packing of rounded copies of some of the input objects increases the required bin count by at most a small fraction of the optimal count, plus an instance-independent additive constant.
Then, we show how to compute an optimal packing with this arrangement in polynomial time.
More concretely, this arrangement consists of a hierarchy of square containers, each of which may contain copies of comparable diameter objects and smaller square containers reserved for smaller copies.

To construct such packings, we first partition the input objects by diameter intervals, then select those in periodically spaced intervals for separate packing.
We call these objects the \emph{separator set}.
The remaining objects are then partitioned into \emph{levels} separated by diameter gaps.
The top level will be packed directly in unit bins, and each lower level has its own square-container size.
At each level, object diameters are bounded from below relative to the container side length, and at lower levels they are also bounded from above by an accuracy-dependent fraction of that length.
The grid assumption then bounds an object's area from below by a class-dependent multiple of its squared diameter, so the diameter lower bound limits how many objects of one level fit in its square, while the upper bound controls the probability that an object crosses a randomly shifted grid line.
Finally, an averaging argument ensures that, for at least one separator choice, the computed separate packing uses a small fraction of the optimal number of bins for the whole instance, plus an additive constant.

We then apply representative rounding to the levels and call the resulting multiset of copies the \emph{rounded supply}.
Shapes such as disks and squares have a natural size order compatible with containment, whereas diameter alone gives no such order for general polygons.
Under bounded translation-containment width, we construct the rounding by applying one-dimensional linear grouping~\cite{VegaLueker81} along containment chains, following the ordered-replacement viewpoint of Eisenbrand, Kakimura, Rothvo{\ss}, and Sanit\`a~\cite{EKRS11}.
We divide each chain into consecutive segments of a common cardinality, except possibly the last, and select its first segment for separate packing.
Each later segment forms a containment group and is replaced by copies of its maximum.
In any reference solution for the input objects, these copies fit inside distinct placements of objects from the preceding segment, so the reference bin count is preserved for the rounded supply.
Conversely, every input object retained for rounding fits within its group's maximum, so after packing the rounded copies, we can substitute the input objects for the rounded copies.

Returning to the general framework, representative roundability ensures that the rounded supply for each separator choice fits into at most $\OPT$ bins.
For the analysis, fix a packing of that supply in $m$ bins.
We then convert this packing into a \emph{structured packing}, having the mentioned simplified arrangement organized by a square hierarchy, using an arbitrarily small fraction of $m$ additional bins plus an instance-independent additive constant.
To do so, we draw nested square grids with a common uniformly random shift in every bin and delete lower-level copies in three steps.
The starting packing and these grids are analysis-only witnesses.

First, we delete copies cut by the grid lines of their own level.
Because their diameter upper bound makes the crossing probability small, area-weighted averaging gives a common shift for which the deleted copies have total area at most a prescribed fraction of $m$.
This is the averaging principle of Arora's shifted-dissection crossing argument~\cite[\S2.2]{Aro98}, with object area in place of tour length.

Second, we remove all surviving lower-level copies lying in regions of the coarsest grid cells which are clipped by bin boundaries.
The coarsest lower-level grid is fine enough that these boundary strips occupy at most a prescribed fraction of the total bin area.
Because the removed copies are interior-disjoint, the same area bound applies.

For the third deletion, note that a square reserved for deeper copies may overlap a larger copy, even though the copies themselves have disjoint interiors.
Accordingly, we delete the deeper copies in each such square and stop processing the square's descendants.
The total area of these deleted copies can be charged to the portions of these squares outside the larger copy, which lie in the excess of its grid-cell cover.
Since the scales are separated, the grid for the smaller squares is sufficiently fine relative to the larger copy, so the grid assumption bounds this excess by a prescribed fraction of that copy's area.
Thus, processing from larger to smaller regions in this way charges each deleted copy at most once, and the interior-disjoint larger copies have total area at most $m$.
The surviving copies therefore admit a structured packing, and the three deletion steps together remove copies of total area at most a prescribed fraction of $m$.

To treat the deleted copies, we first fit each into a square container with side length given by its level, then nest these containers in additional bins.
The area lower bound from the grid assumption controls the total squared diameter and bounds the number of additional bins.
These steps construct a structured packing of the entire rounded supply and, together, bound the structured optimum.
The algorithm computes neither this witness nor the grid shift or deletions.

We then address how to compute an optimal structured packing for the entire rounded supply.
At each level, we pack its rounded copies into square regions and, when needed, reserve smaller squares, disjoint from the copies and one another, for the next level.
The top-level regions are unit bins, and all regions at a given level have the same size.

For each level, we enumerate the possible counts of rounded copies from each containment group in one region.
These counts form a \emph{configuration}.
The bounds on local occupancy and the number of containment groups bound the number of feasible configurations in terms of the class and accuracy.
Using exact translational containment~\cite{Mil97}, we determine which configurations are feasible and, for each feasible one, compute the maximum number of smaller reserved squares, called \emph{child squares}, that fit jointly with its copies.
We store rational placements of both the copies and child squares, with no grid-alignment requirement.
As a result, each local query contains only a bounded number of movable objects for fixed class and accuracy, so computations take polynomial time even when boundary descriptions are part of the input.

For each separator choice, we formulate an integer program to decide how many times each feasible configuration is used, minimizing the number of unit bins while exactly covering the rounded supply.
The constraints ensure that each level reserves enough child squares for all regions at the next level.
Only total child capacity matters, since all regions at the next level have the same size and their entire packings can be translated into any reserved square of that size.
Each level contributes a bounded number of variables for fixed class and accuracy, each variable has a polynomially bounded range, and every constraint involves one level or two consecutive ones.
With the minimum-cost extension of Jansen and Kratsch's bounded-domain feasibility dynamic program~\cite{JK15}, we can therefore solve this integer program in polynomial time.

From the optimal counts, we assemble the stored local realizations by translating whole child packings into their assigned reserved squares.
We then substitute each containment group's input objects bijectively into its rounded-copy placements, preserving feasibility by containment.
Next we combine this packing with the two separate packings of the separator set and the input objects selected at rounding.
We evaluate every separator choice and return a resulting solution with the best cost.
Finally, we prove that the solution uses at most $1+\eps$ times the optimum number of bins, plus a constant depending only on the class and accuracy, thus establishing an \aptas{}.

\paragraph{Organization of the text.}

In Section~\ref{sec:preliminaries}, we formalize the geometric assumptions and describe a routine for containment comparisons and packing prescribed objects in one container.
In Section~\ref{sec:abstract-geometry}, we partition the instance into levels, and in Section~\ref{sec:nesting-realization}, we establish representative roundability under bounded translation-containment width.
Next, in Section~\ref{sec:transfer}, we show that a structured comparison packing with all desired properties exists.
In Section~\ref{sec:local-configurations}, we compute a structured packing of the rounded supply into the minimum number of bins and demonstrate the overall guarantees.
Finally, Section~\ref{sec:conclusion} extends the framework to unit-cube bins in every fixed dimension.

\clearpage
\section{Preliminaries}\label{sec:preliminaries}

In this work, a geometric bin packing instance is a labeled multiset $I=\{P_1, \ldots, P_n\}$ of \emph{objects} which are compact, connected, non-degenerate polygonal regions, each specified by a non-self-intersecting rational outer boundary cycle and finitely many non-self-intersecting rational inner boundary cycles, with each cycle having at least three vertices.
The closed regions bounded by the inner cycles are pairwise disjoint and lie strictly inside the outer cycle.
In particular, holes are permitted and nested holes are excluded.
We fix a class $\calP$ of such objects and consider instances drawn from $\calP$.
We allow translations only.
Let $L\ge2$ be the total binary input length.

A \emph{placement} of an object $P_i \in I$ is a translated copy of it.
A \emph{packing} of $I$ is a placement of all of its objects into unit-square bins such that the objects have pairwise disjoint interiors with boundary contact allowed.
We assume, without loss of generality, that every input object fits into $[0,1]^2$, and write $\OPT(I)$ for the minimum number of bins required to pack $I$ and $\area(I):=\sum_{P_i\in I}\area(P_i)$.
We are interested in results of the following type.

\begin{definition}[APTAS]
  \label{def:ptas}
  An \emph{asymptotic polynomial-time approximation scheme} (APTAS) for $\calP$ is a family of algorithms $\{A_\eps\}_{0<\eps<1}$ such that $A_\eps$ runs in time $L^{g_{\calP}(1/\eps)}$ and returns a packing whose bin count $A_\eps(I)$ satisfies $A_\eps(I)\le(1+\eps)\OPT(I)+\delta_{\eps,\calP}$ for every instance $I$ drawn from $\calP$, where $g_{\calP}$ is computable and $\delta_{\eps,\calP} \in \mathbb Z_{\ge 0}$ is independent of $n$, $L$, and the instance.
\end{definition}

The translation-containment relation for $I$ is
\begin{equation}
\label{eq:containment} P_i\preceq P_j\quad\Longleftrightarrow\quad\exists v\in\mathbb R^2\ (P_i+v\subseteq P_j).
\end{equation}
Note that this relation is a preorder.

Fix an object $P$.
Let $D(P):=\max_{x,y\in P}\lVert x-y\rVert_2$ be its diameter, and let $b(P)$ be the larger of the horizontal and vertical spans of $P$.
We always assume normalized objects in the input where their minimum coordinates are zero.
Define $B(P):=[0,b(P)]^2$ as a canonical bounding square of $P$.
The diameter and span satisfy:
\begin{equation}
\label{eq:bdiam} b(P)\le D(P)\le\sqrt2\cdot b(P).
\end{equation}
Throughout, we distinguish input objects by their labels, even when their normalized geometries are equal.
Moreover, a set of objects means a set of labeled occurrences, so repeated geometries retain their multiplicities.

\begin{definition}[Square grid, phase, and grid-cell cover]
  \label{def:grid}
  The \emph{square grid} $\mathcal G(h,\tau)$ has cell side length $h>0$ and lines $x=\tau_x+zh$, $y=\tau_y+z'h$ for $z,z'\in\mathbb Z$, with phase $\tau$ understood modulo $h$.
  We define the \emph{grid-cell cover of $P$ by $\mathcal G(h,\tau)$}, denoted $V_{h,\tau}(P)$, as the union of the closures of cells $C$ of $\mathcal G(h,\tau)$ for which $\operatorname{int}(C)\cap\operatorname{int}(P)\ne\varnothing$.
  We write $V_h(P)$ when a phase is fixed by context or an assertion holds uniformly over phases.
\end{definition}

Our framework begins with a preprocessing step that selects objects from periodically spaced diameter intervals for separate packing and divides the remainder into sets of comparable diameter.
Within each set, a rounding operation selects additional objects for separate packing, partitioning the rest into \emph{containment groups}, then replaces objects in a group with copies of the group maximum under $\preceq$.
The rounding assumption bounds the number of chosen maxima per set independently of its cardinality, and it ensures that the resulting copies admit a packing in at most $\OPT(I)$ bins.
The grid assumption controls the excess area introduced by sufficiently fine grid-cell covers.
We now formally state both assumptions, followed by the algorithmic guarantee they support.

\paragraph{Grid assumption.}

For a fixed $\calP$, the grid assumption requires a function that maps each target relative error to a bound on the allowed square grid cell side length, as a fraction of an object's bounding-square side length.
This same relative bound must then control the excess covered area uniformly over all members, placements, and grid phases.
Formally, we require the following.

\begin{definition}[Uniform rasterizability]
  \label{def:gridregular}
  A class of objects $\calP$ is \emph{uniformly rasterizable} if there exists a polynomial-time computable function $\gridrate:(0,1)\cap\mathbb Q\to(0,1]\cap\mathbb Q$ with the following property.
  For every rational $0<\eta<1$, every placement $P$ of an object of $\calP$ and every square grid $\mathcal G(h,\tau)$ of arbitrary phase with $h\le\gridrate(\eta)\cdot b(P)$, the grid-cell cover of $P$ by $\mathcal G(h,\tau)$ satisfies
  \begin{equation}
  \label{eq:outer-abstract} \area\bigl(V_h(P)\setminus P\bigr) \le \eta \cdot \area(P).
  \end{equation}
\end{definition}

We call $\gridrate$ the \emph{grid-resolution function} of $\calP$.
The motivation comes from classical metric neighborhoods in topology and geometry: a Euclidean $\delta$-neighborhood, or parallel set, enlarges a set by a ball of radius $\delta$ through Minkowski sum, and its volume growth underlies notions such as Minkowski content and growth~\cite[Chap.~3]{Fal03}.
Instead, we require a uniform bound over the class on added area relative to an object's area.
Our definition is arguably more applicable in computational geometry contexts, as it provides a fine-grained notion of boundary complexity relative to polygon size. 
Besides, uniform rasterizability is equivalent, up to constant rescaling, to a uniform relative outer Minkowski growth concept.
The proof of this equivalence is given in Appendix~\ref{app:minkowski-proofs}.

\begin{proposition}
  \label{prop:grid-bulk}
  If $\calP$ is uniformly rasterizable, it satisfies the following uniform bulk inequality
  \begin{equation}
  \label{eq:lambda-grid} \area(P)\ge \lambda\cdot D(P)^2 \qquad(P\in\calP), \qquad \lambda:=\nicefrac{1}{6}\min\left\{\gridrate(\nicefrac{1}{2}),\nicefrac{1}{2}\right\}>0.
  \end{equation}
\end{proposition}

The definition of $\lambda$ uses the rasterizability guarantee at the fixed error $\nicefrac{1}{2}$.
The bulk inequality can be used to connect uniform rasterizability to classical fatness.
In Appendix~\ref{app:concrete}, we prove that uniform rasterizability and uniform fatness are equivalent for classes of convex polygons and incomparable for general non-convex classes.

This inequality serves two purposes in the framework, and its proof is deferred to Appendix~\ref{app:grid-bulk}.
First, it converts bounds on the total area of separately packed objects into bounds on their total squared diameter, which controls the number of bins needed for that packing.
Second, it bounds the number of objects of comparable diameter that fit in a square bin of fixed side length.

\paragraph{Rounding assumption.}

Our representative-rounding interface adapts Eisenbrand, Kakimura, Rothvo{\ss}, and Sanità's~\cite{EKRS11} abstract ordered-replacement viewpoint to geometric bin packing.
In their framework, smaller elements can replace larger ones without destroying feasibility.

Accordingly, in Section~\ref{sec:nesting-realization}, we use containment chains under the relation $\preceq$ to realize the rounding procedure on sets of objects with comparable diameter.
We apply rounding separately to each such set.
For each application, we fix a square side length and impose a lower bound on diameter relative to this length to control occupancy.
Except for the largest objects, we require diameters to be at most a fixed fraction of the associated square side length.
Next, we formalize these ideas.

\begin{definition}[Level-separated partition]
  \label{def:separated}
  Fix a dyadic parameter $\alpha=2^{-K}$ for a positive integer $K$ and constants $0<\beta,\gamma<1$.
  Let $I$ be a finite multiset of objects.
  A partition $(S_0,\ldots,S_J)$ of $I$, together with positive reals $s_0,\ldots,s_J\in(0,1]$, is an \emph{$(\alpha,\beta,\gamma)$-level-separated partition} if $s_0=1$ and
  \[
    \beta\cdot s_j < D(P) \quad(P\in S_j), \qquad D(P)\le \alpha\cdot s_j\quad(P\in S_j,\ j\ge1), \qquad s_{j+1}=\gamma\cdot s_j\quad(1\le j<J).
  \]

  We call each $j = 0, 1, \dots, J$ and its corresponding set $S_j$ a \emph{level}, with level zero being the \emph{top level} and levels one onward, the \emph{lower levels}.
\end{definition}

In Section~\ref{sec:abstract-geometry}, we construct a level-separated partition by selecting objects from periodically spaced diameter intervals for separate packing.
Our goal is to obtain a simplified packing of such objects that can later be computed in polynomial time.
To do so, we adopt a rounding interface that bounds the number of containment groups per level and preserves packability for the remaining rounded objects after selecting a few original objects for separate packing.
The first guarantee bounds the number of shapes at each level.
The second ensures that the rounded objects can be packed in at most $\OPT(I)$ bins.

\begin{definition}[Uniform representative roundability]
  \label{def:roundable}
  $\calP$ is \emph{uniformly representative-roundable} if, for every dyadic $\alpha$, $0 < \beta, \gamma<1$, and $0<\eta<1$, given any $(\alpha,\beta,\gamma)$-level-separated partition of an input $I\subseteq\calP$, we can in polynomial time select a set $E_{\mathrm{grp}}$ for separate packing, partition each $S_j\setminus E_{\mathrm{grp}}$ into at most $\Rrep$ containment groups $\Sclass_{j,u}$, each with a chosen local maximum $\widehat P_{j,u}\in\Sclass_{j,u}$ under $\preceq$, and compute a packing of $E_{\mathrm{grp}}$.

  Moreover, if $I$ can be packed in $m$ bins, replacing each object in a containment group by a copy of its chosen maximum yields a multiset that can be packed in $m$ bins, and the computed packing of $E_{\mathrm{grp}}$ uses at most $\eta \cdot m+\delta_{\mathrm{grp}}$ bins.
  Furthermore, $\Rrep$ and $\delta_{\mathrm{grp}}$ may depend only on $\alpha$, $\beta$, $\gamma$, and $\eta$.
  For rational $\alpha,\beta,\gamma,\eta$, we can perform these computations using a single algorithm that takes these parameters as input.
  Computable functions of the parameters bound $\Rrep$, the multiplicative constant, and the exponent in its polynomial running time.
\end{definition}

Let $\widehat S_j$ be the multiset including a copy of $\widehat P_{j,u}$ for each input object $P\in\Sclass_{j,u}$, over all level-$j$ containment groups.
We call each such copy a \emph{rounded copy} and $\widehat S:=\dot{\bigcup}_{j}\widehat S_j$ the \emph{rounded supply}.
By the preservation-of-packability guarantee in Definition~\ref{def:roundable}, we have $\OPT(\widehat S) \le \OPT(I)$.
Moreover, by Proposition~\ref{prop:grid-bulk} and the lower bound on diameter, each level-$j$ object has area greater than $\lambda \cdot \beta^2 \cdot s_j^2$.
Thus, a square of side $s_j$ contains at most $\Kloc$ pairwise interior-disjoint objects from level $j$, where
\begin{equation}
\label{eq:K}
\Kloc:=\left\lceil\frac{1}{\lambda\cdot\beta^2}\right\rceil.
\end{equation}
Together with the bound on the number of chosen maxima, this occupancy bound implies that the number of valid, distinct multiplicity vectors is limited, independent of level and input cardinality. \medskip

Finally, both the grid and rounding assumptions yield our main result.

\begin{RoundedBox}{\textsc{Main result}}
  \begin{theorem}
    \label{thm:abstract-main}
    If $\calP$ is uniformly rasterizable and uniformly representative-roundable, bin packing for $\calP$ admits an APTAS without bin augmentation, with rational translation coordinates.
  \end{theorem}
\end{RoundedBox}

We illustrate Theorem~\ref{thm:abstract-main} with two concrete polygon classes by verifying uniform rasterizability and uniform representative roundability.
First, recall that the \emph{width} of a partial order is the maximum cardinality of a set of pairwise incomparable elements.
For a uniformly rasterizable class, a class-dependent bound on translation-containment relation width implies uniform representative roundability, as proved in Section~\ref{sec:nesting-realization}.
Accordingly, we have the following.

\begin{definition}[Translation-containment width]
  \label{def:containment-width}
  If every instance $I\subseteq \calP$ has an underlying translation-containment relation $\preceq$ of width at most $d$, for a positive integer $d$, we say that $\calP$ has \emph{translation-containment width at most $d$}.
\end{definition}

In Appendix~\ref{app:concrete}, we verify these hypotheses for some concrete classes of objects and prove the following two corollaries.

\begin{corollary}
  \label{cor:homothetic-star-ptas}
  If $\calP$ is formed by homothetic copies of a finite set of star-shaped objects, bin packing for $\calP$ admits an APTAS without bin augmentation, with rational translation coordinates.
\end{corollary}

\label{def:fatness}
A polygon $P$ is $f$-fat if $r^{\mathrm{out}}(P)/r^{\mathrm{in}}(P)\le f$, where these are its circumradius and inradius.
In particular, we say that $\calP$ is uniformly $f$-fat if every member is $f$-fat.
For convex classes, uniform fatness in turn supplies uniform rasterizability, and we obtain the following result.

\begin{corollary}
  \label{cor:concrete}
  If $\calP$ is a uniformly $f$-fat class of convex polygons, bin packing for $\calP$ admits an APTAS without bin augmentation, with rational translation coordinates, whenever $\calP$ has bounded translation-containment width.
\end{corollary}

\subsection{Exact packing in one container}

The framework needs to compute exact containment for $\preceq$ and to verify whether a given small set of objects can be packed in a square.
To do so, we use the classical contact-enumeration method of Daniels and Milenkovic for packing prescribed objects in a polygonal container by translation~\cite{DM97,Mil97}.
The following local result requires neither uniform rasterizability nor representative roundability.

\begin{lemma}
  \label{lem:local-solver}
  Given objects $P_1, \ldots, P_k$ and a container $Q$ with the rational polygonal boundary representation specified at the start of this section, let $L_{\mathrm{loc}}$ denote their total binary encoding length.
  We can decide whether there is a packing of $P_1, \ldots, P_k$ in $Q$ in time $F_{\mathrm{loc}}(k) \cdot L_{\mathrm{loc}}^{O(k+1)}$ and, in the affirmative, compute rational translation vectors realizing such a packing, with encoding length polynomial in $L_{\mathrm{loc}}$.
  Here $F_{\mathrm{loc}}$ is computable and non-decreasing.
\end{lemma}

To apply the contact-enumeration method, we first triangulate the objects and a region outside the container.
Precisely, we fix a strictly enclosing rational rectangle $R$ for $Q$ and triangulate $\overline{R\setminus Q}$, respecting all boundary cycles.
This covers both the holes of $Q$ and the region between its outer boundary and $R$.
Then, an object is entirely contained in $Q$ if and only if all its triangles lie within $R$ and are interior-disjoint from all fixed complement triangles.
Moreover, two objects are interior-disjoint if and only if their respective triangles are interior-disjoint.
Therefore, we reduce the problem of checking whether given translation candidates ensure a feasible packing to checking interior-disjointness of triangles.
Two triangles have disjoint interiors precisely when they lie in opposite closed half-planes bounded by a line parallel to an edge of either triangle.
The objects cannot rotate, so their triangle edges have fixed directions.
The containment and interior-disjointness constraints can therefore be expressed as a conjunction of constant-size disjunctions of weak rational linear inequalities.
The number of inequalities and the lengths of their coefficient encodings are polynomial in $L_{\mathrm{loc}}$.

Let $N=L_{\mathrm{loc}}^{O(1)}$ be the number of boundary hyperplanes of the inequalities that ensure the containment and interior-disjointness constraints, including those defined by bounds that keep each object inside the enclosing rectangle.
Each selection of one separating inequality for every pair of triangles that must be interior disjoint, together with the bounds from the enclosing rectangle, defines a possibly empty bounded rational polytope.
The set of feasible translation tuples is hence the union of these polytopes.
Every nonempty such polytope has a vertex determined by $2k$ linearly independent active constraint hyperplanes, even when it has empty interior.
Accordingly, we enumerate subsets of $2k$ hyperplanes with linearly independent normals and solve their defining equations to obtain candidate translation tuples.
We then test each of the at most $N^{2k}$ candidates against the containment and interior-disjointness constraints.
Exact rational linear algebra gives the stated running time and polynomial encoding length.

For containment comparisons, we use $k=1$, while the local packing queries will have $k$ determined only by $\calP$ and $\varepsilon$.
Because boundary complexity and coordinate encoding remain part of $L_{\mathrm{loc}}$, these calls take polynomial time for fixed $\varepsilon$.

\section{Geometric Preprocessing}\label{sec:abstract-geometry}

For fixed rational $0<\eps<1$, we construct a level-separated partition as in Definition~\ref{def:separated} and bound the number of bins needed for separately packed objects.
A smaller positive rational accuracy suffices for any real target accuracy, so this restriction preserves the APTAS guarantee.

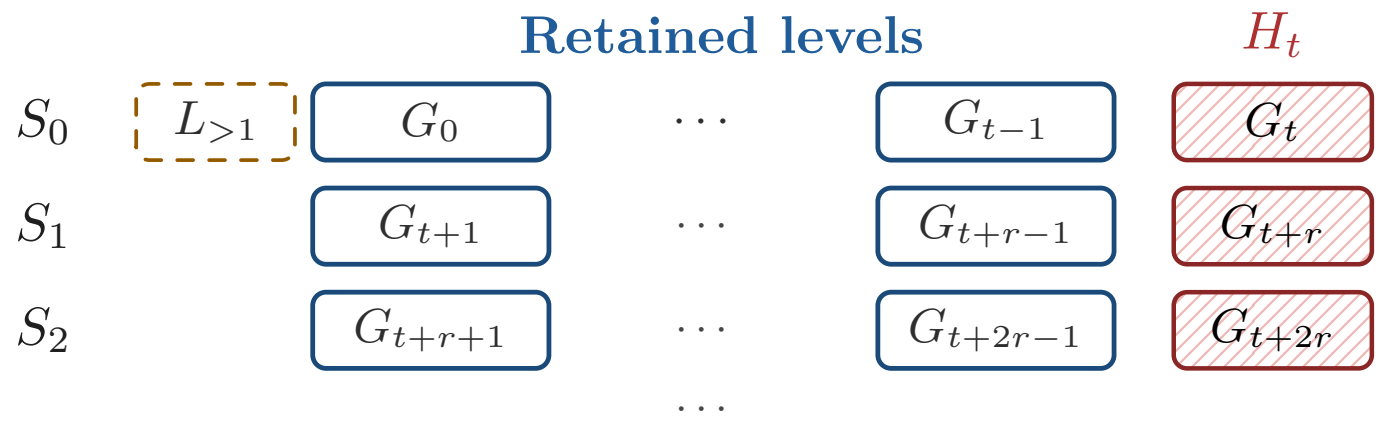
\begin{figure}[H]
  \centering
  \begin{tikzpicture}[x=0.84cm,y=0.84cm,font=\small,line cap=round,line join=round]
    \node[figtitle,text=aptasblue] at (5.96,3.35) {Retained levels};
    \node[figtitle,text=aptasred] at (10.30,3.35) {$H_t$};

    \node[figtitle] at (0.62,2.66) {$S_0$};
    \node[figtitle] at (0.62,1.84) {$S_1$};
    \node[figtitle] at (0.62,1.02) {$S_2$};

    \path[rootband] (1.35,2.36) rectangle (2.60,2.96);
    \node[figlabel] at (1.98,2.66) {$L_{>1}$};
    \foreach \x in {3.67,8.12} {
        \path[retainedband]
        ({\x-0.93},2.36) rectangle ({\x+0.93},2.96);
      }
    \foreach \x/\entry in {3.67/G_0,5.83/\cdots,8.12/G_{t-1}} {
    \node[figlabel] at (\x,2.66) {$\entry$};
    }

    \foreach \y/\bandfirst/\bandlast in {1.84/G_{t+1}/G_{t+r-1},1.02/G_{t+r+1}/G_{t+2r-1}} {
    \path[retainedband]
    (2.74,{\y-0.30}) rectangle (4.60,{\y+0.30});
    \path[retainedband]
    (7.19,{\y-0.30}) rectangle (9.05,{\y+0.30});
    \node[figlabel] at (3.67,\y) {$\bandfirst$};
    \node[fignote] at (5.83,\y) {$\cdots$};
    \node[figlabel] at (8.12,\y) {$\bandlast$};
    }
    \node[fignote] at (5.83,0.4) {$\dots$};

    \foreach \y/\band in {2.66/G_t,1.84/G_{t+r},1.02/G_{t+2r}} {
    \path[separatorband]
    (9.52,{\y-0.30}) rectangle (11.08,{\y+0.30});
    \node[figlabel,text=black] at (10.30,\y) {$\band$};
    }

  \end{tikzpicture}
  \caption{
    Level-separated partition by separator residue.
    The presentation adapts the table organization of Miyazawa--Pedrosa--Schouery--Sviridenko--Wakabayashi~\cite{MPSSW14}.
  }
  \label{fig:scale-separation}
\end{figure}

For a dyadic $0<\alpha<1$ chosen as in Lemma~\ref{lem:preprocessing}, we partition the objects of diameter at most $1$ by diameter:
$G_i:=\{P\in I\mid\alpha^{2i}\ge D(P)>\alpha^{2(i+1)}\}$ for all $i \ge 0$.
We choose the separator period $r:=\left\lceil\frac{32}{\lambda\cdot\eps}\right\rceil+2$ and, for each \emph{separator residue} $t\in\{1,\ldots,r-1\}$, define a \emph{separator} set
\[
  H_t:=\bigcup_{i \equiv t \pmod r}G_i.
\]
We omit residue $0$ to keep $G_0$ at the top level.
Given $H_t$, the sets $G_i$ retained between successive diameter gaps form sequences of $r-1$ consecutive, possibly empty sets.
Let these sequences define the lower-level sets $S_j$ for all $j \ge 1$, that is, the retained level $S_j$ is the union of $G_{t+(j-1)r+1}$ through $G_{t+jr-1}$.
Likewise, the top level $S_0$ contains the prefix before the first gap and the set $L_{>1}:=\{P\in I\mid D(P)>1\}$.
Figure~\ref{fig:scale-separation} illustrates this construction.
We associate these levels with the reals $s_0:=1$ and $s_j:=\alpha^{2(t+(j-1)r)+1}$ for $j\ge1$.
Let $\beta:=\alpha^{2r-1}$, and $\gamma:=\alpha^{2r}$, hence $s_{j+1}=\gamma\cdot s_j$ and $s_1 \le \alpha^3$.
Furthermore, every object in $H_t$ has diameter at most $\alpha^2$.
Finally, the diameter bounds give $\beta\cdot s_j<D(P)\le\alpha\cdot s_j$ for every $P\in S_j$, $j\ge 1$.
Therefore, these $S_i$'s, together with their associated $s_j$'s, form an $(\alpha, \beta, \gamma)$-level-separated partition of $I\setminus H_t$.

The diameter lower bound controls the number of objects of $S_j$ that fit in a square of side $s_j$, as stated in  \eqref{eq:K}.
For $j\ge1$, the diameter upper bound, in turn, limits the probability that an object's interior crosses a line of a randomly shifted square grid of cell side $s_j$.
These facts later support packing each rounded copy of $S_j$ from $\widehat S_j$ separately into squares of side $s_j$.
Because $\gamma=\alpha\cdot \beta$, then $s_{j+1} \leq \alpha\cdot  D(P)$ for every object at a nonterminal lower level.
Ultimately, for a target relative excess-area error $\eta$, choosing $\alpha$ so that $\alpha\cdot  D(P)\le \gridrate(\eta)\cdot b(P)$ lets rasterizability control the grid-cell cover excess at that scale.
At level $0$, the separate bound $D(P)>\alpha^{2t}>\beta\cdot s_0 = \beta$ gives the same relation $s_1<\alpha\cdot D(P)$.
Because $\alpha$ is chosen as a reciprocal power of two, every ratio $\nicefrac{s_j}{s_{j+1}}$ between consecutive square side lengths is an integer.

In the following lemma, we summarize the level-separated partition constructed above and state the required parameters.
The proof details are presented in Appendix~\ref{app:scale-construction}.

\begin{lemma}
  \label{lem:preprocessing}
  For a uniformly rasterizable class $\calP$ and fixed rational $0<\eps<1$, we can in polynomial time compute parameters $r,\alpha,\beta,\gamma$ depending only on $\calP$ and $\eps$, with dyadic $\alpha\le \nicefrac{1}{16}$, and $\gamma\le \nicefrac{1}{2}$.
  Given an input instance $\emptyset \ne I \subseteq \calP$, for each $t\in\{1, \ldots, r-1\}$, we can construct in polynomial time a separator set $H_t$ and an $(\alpha, \beta, \gamma)$-level-separated partition of $I\setminus H_t$ comprising $(S_0, \ldots, S_J)$ and $s_0,\ldots,s_J\in(0,1]$.
  Empty levels are retained, and $S_J\ne\emptyset$ whenever $I\setminus H_t\ne\emptyset$.
  If $I\setminus H_t=\emptyset$, we take $J=0$ and $S_0=\emptyset$.
  For $J\ge1$, all ratios $\nicefrac{s_j}{s_{j+1}}$ are integers bounded by $\Mgrid:=\nicefrac{1}{\gamma}$, and for every $j<J$ and $P\in S_j$,
  \begin{equation}
    \label{eq:child-resolution}
    s_{j+1}<\alpha \cdot D(P)\le\gridrate(\nicefrac{\lambda\eps}{64}) \cdot b(P).
  \end{equation}
  Moreover, $J=O(L)$, and all scales therefore have polynomial encoding length for fixed $\eps$.
\end{lemma}

We still need to pack the separator set $H_t$.
To do so, we use the Next-Fit Decreasing Height (NFDH) algorithm over the objects' bounding squares.
In the following lemma, we state the residue bound, which follows from an averaging argument with the square-packing bound in~\cite{CGJT80}, and it is proved in Appendix~\ref{app:separator}.

\begin{lemma}
  \label{lem:medium}
  At least one separator residue $t\in\{1,\ldots,r-1\}$ satisfies
  \begin{equation}
    \label{eq:cheap-residue}
    \area(H_t)\le\frac{\OPT(I)}{r-1}.
  \end{equation}

  Hence, for every separator residue satisfying this bound, NFDH packs the bounding squares of $H_t$ in at most
  \begin{equation}
  \label{eq:medium-cost} \frac{4}{\lambda\cdot(r-1)}\OPT(I)+2 \le\frac{\eps}{8}\OPT(I)+2
  \end{equation}
  unit-square bins.
\end{lemma}
Our framework evaluates every residue and returns the packing with the fewest bins, so it does not need to identify the residue the lemma guarantees in advance.

\section{Representative Roundability from Translation Containment}
\label{sec:nesting-realization}

For each separator set $H_t$, preprocessing yields a level-separated partition of the retained input objects $I \setminus H_t$.
We now verify that the assumptions of Definition~\ref{def:roundable} hold under bounded translation-containment width.
In Appendix~\ref{app:width-rounding}, we give complete proofs of the sketch below.

\begin{lemma}
  \label{lem:width-roundable}
  If $\calP$ is uniformly rasterizable and has translation-containment width at most $d$, then it is uniformly representative-roundable.
\end{lemma}

We note that this width bound suffices for the construction below, although the general representative-rounding interface does not assume it. 
Using Dilworth's theorem~\cite{Dil50}, we first partition each level into at most $d$ containment chains, and we compute the required containment comparisons via Lemma~\ref{lem:local-solver}.
We then round each chain separately, adapting the linear grouping idea for one-dimensional bin packing~\cite{VegaLueker81}. 

Fix $0<\eta<1$.
Within each level $S_j$, we order the objects in each chain non-increasingly under $\preceq$, then divide each chain into consecutive segments $T_1, \ldots, T_\ell$ of a common size for that level, allowing a shorter final segment.
We include each first segment $T_1$ in $E_{\mathrm{grp}}$.
We then index the later segments across all chains at level $j$ by $u$, write $\Sclass_{j,u}$ for the segment with index $u$, and choose its first member $\widehat P_{j,u}$ as a local maximum.
For each input object $P\in\Sclass_{j,u}$, we add to $\widehat S_j$ a copy of $\widehat P_{j,u}$ carrying the label of $P$.
The chain order gives $P\preceq\widehat P_{j,u}$ for every $P\in\Sclass_{j,u}$.
This, in turn, allows arbitrary bijective substitution within each containment group during reconstruction.
Accordingly, the segment size balances the number of containment groups against the number of input objects packed separately.
Uniform rasterizability enters through the bulk bound of Proposition~\ref{prop:grid-bulk} and the occupancy bound~\eqref{eq:K}, which together determine the choice of segment size.
Appendix~\ref{app:width-rounding} gives a suitable choice, the bounds on $\Rrep$ and $\delta_{\mathrm{grp}}$, and the explicit packing of $E_{\mathrm{grp}}$.

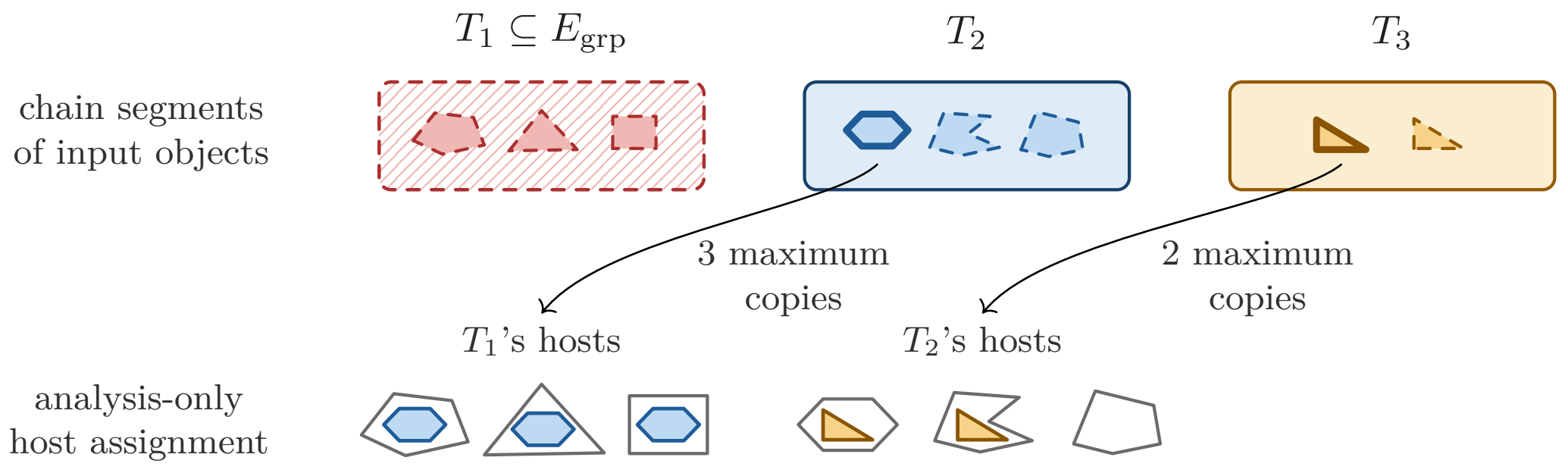
\begin{figure}[!t]
  \centering
  \begin{tikzpicture}[
      x=1.35cm,y=1cm,font=\small,line cap=round,line join=round,
      pics/gTwoRepresentative/.style={code={
              \draw[pic actions]
              (-0.20,0)--(-0.10,0.14)--(0.10,0.14)--(0.20,0)--(0.10,-0.14)--(-0.10,-0.14)--cycle;
            }},
      pics/gThreeRepresentative/.style={code={
              \draw[pic actions]
              (-0.16,-0.13)--(-0.16,0.13)--(0.16,-0.13)--cycle;
            }}
    ]
    \node[figlabel,anchor=east] at (1.34,3.68) {chain segments\\of input objects};
    \node[figlabel,anchor=east] at (1.34,1.15) {analysis-only\\host assignment};

    \foreach \x/\heading in {3.00/{T_1\subseteq E_{\mathrm{grp}}},5.75/{T_2},8.50/{T_3}} {
        \node[figtitle] at (\x,4.55) {$\heading$};
    }

    \path[draw=aptasred,fill=white,pattern=north east lines,pattern color=aptasredlight,
      rounded corners=3pt,line width=0.85pt,dashed] (1.95,3.18) rectangle (4.05,4.12);
    \path[chainboxmarked] (7.45,3.18) rectangle (9.55,4.12);
    \path[draw=aptasred,fill=aptasredlight,line width=0.8pt,dashed]
    (2.17,3.61)--(2.31,3.85)--(2.56,3.81)--(2.63,3.57)--(2.36,3.49)--cycle;
    \path[draw=aptasred,fill=aptasredlight,line width=0.8pt,dashed]
    (2.79,3.52)--(3.00,3.87)--(3.23,3.53)--cycle;
    \path[draw=aptasred,fill=aptasredlight,line width=0.8pt,dashed]
    (3.46,3.54) rectangle (3.74,3.82);
    
    \path[chainboxplain] (4.7,3.18) rectangle (6.8,4.12);
    \pic[draw=aptasblue,fill=aptasbluelight,line width=1.7pt]
    at (5.17,3.70) {gTwoRepresentative};
    \path[draw=aptasblue,fill=aptasbluelight,line width=0.8pt,dashed]
    (5.51,3.55)--(5.60,3.85)--(5.90,3.82)--(5.76,3.67)--(5.96,3.55)--(5.72,3.48)--cycle;
    \path[draw=aptasblue,fill=aptasbluelight,line width=0.8pt,dashed]
    (6.10,3.53)--(6.20,3.86)--(6.47,3.77)--(6.50,3.53)--(6.28,3.47)--cycle;

    \path[draw=aptasamber!75!black,fill=aptasamberlight,line width=1.7pt]
    (8.01,3.53)--(8.01,3.79)--(8.33,3.53)--cycle;
    \path[draw=aptasamber!80!black,fill=aptasamberlight,line width=0.8pt,dashed]
    (8.64,3.54)--(8.64,3.80)--(8.96,3.54)--cycle;

    \draw[->,draw=black,line width=0.5pt]
    (5.17,3.4) .. controls (4.86,3.02) and (3.39,2.78) .. (3.00,2.10);
    \draw[->,draw=black,line width=0.5pt]
    (8.17,3.4) .. controls (7.82,3.02) and (6.25,2.78) .. (5.85,2.10);
    \node[figlabel,fill=white,inner sep=1.5pt] at (4.63,2.4) {3 maximum\\copies};
    \node[figlabel,fill=white,inner sep=1.5pt] at (7.63,2.4) {2 maximum\\copies};

    \node[figlabel] at (3.00,1.86) {$T_1$'s hosts};
    \begin{scope}[shift={(2.18,1.13)}]
      \draw[draw=aptasgray,line width=0.8pt]
      (-0.345,-0.09)--(-0.135,0.27)--(0.24,0.21)--(0.345,-0.15)--(-0.06,-0.27)--cycle;
    \end{scope}
    \begin{scope}[shift={(3.00,1.13)},scale=1.10]
      \draw[draw=aptasgray,line width=0.8pt]
      (-0.336,-0.24)--(0,0.32)--(0.368,-0.224)--cycle;
    \end{scope}
    \begin{scope}[shift={(3.82,1.13)}]
      \draw[draw=aptasgray,line width=0.8pt]
      (-0.252,-0.252)--(-0.252,0.252)--(0.252,0.252)--(0.252,-0.252)--cycle;
    \end{scope}
    \foreach \x in {2.18,3.82} {
        \pic[draw=aptasblue,fill=aptasbluelight,line width=0.9pt]
        at (\x,1.13) {gTwoRepresentative};
    }
    \pic[draw=aptasblue,fill=aptasbluelight,line width=0.9pt]
    at (3.01,1.09) {gTwoRepresentative};

    \node[figlabel] at (5.85,1.86) {$T_2$'s hosts};
    \begin{scope}[shift={(4.98,1.13)}]
      \draw[draw=aptasgray,line width=0.8pt]
      (-0.32,0)--(-0.16,0.224)--(0.16,0.224)--(0.32,0)--(0.16,-0.224)--(-0.16,-0.224)--cycle;
    \end{scope}
    \begin{scope}[shift={(5.85,1.13)}]
      \draw[draw=aptasgray,line width=0.8pt]
      (-0.308,-0.14)--(-0.182,0.28)--(0.238,0.238)--(0.042,0.028)--(0.322,-0.14)--(-0.014,-0.238)--cycle;
    \end{scope}
    \begin{scope}[shift={(6.72,1.13)}]
      \draw[draw=aptasgray,line width=0.8pt]
      (-0.28,-0.168)--(-0.14,0.294)--(0.238,0.168)--(0.28,-0.168)--(-0.028,-0.252)--cycle;
    \end{scope}

    \foreach \x in {4.98,5.85} {
        \path[draw=aptasamber!75!black,fill=aptasamberlight,line width=0.9pt]
        (\x-0.16,0.998)--(\x-0.16,1.258)--(\x+0.16,0.998)--cycle;
      }
  \end{tikzpicture}

  \caption{
    Rounding a containment chain, with analysis-only placement of its rounded copies inside hosts from preceding segments.
    Hatching marks input objects selected for separate packing.
    Later segments $T_2$ and $T_3$ form containment groups whose chosen maxima have thick outlines.
    In the lower row, only colored copies belong to $\widehat S$.
    Gray outlines mark host placements of input objects.
    The drawing is schematic and not to scale.
  }
  \label{fig:chainshift}
\end{figure}

We still need to show that the rounded supply $\widehat S$ can be packed in the same bins as $I \setminus H_t$.
For the analysis, fix any packing of the input objects into $m$ bins.
For every containment group $\Sclass_{j,u}=T_a$, we have $|T_a|\le |T_{a-1}|$.
Moreover, the chain order gives $\widehat P_{j,u}\preceq Q$ for every $Q\in T_{a-1}$.
Thus, for each rounded copy of $\widehat S$ from $\Sclass_{j,u}$, we can assign it to a distinct $Q\in T_{a-1}$ and place the copy inside $Q$'s placement.
These assignments use distinct host labels: different segments use disjoint preceding segments, and different chains and levels are disjoint.
After these assignments, we keep only the rounded copies in this modified packing and delete the remaining objects.
Because each rounded copy lies inside a distinct host placement, the copies have pairwise disjoint interiors and remain packed in the same $m$ bins.
Hence $\OPT(\widehat S)\le m$.
The rounding computation neither receives nor constructs this witness packing.

Figure~\ref{fig:chainshift} shows an example with three segments: the three copies from $T_2$ use the three $T_1$ hosts, and the two copies from $T_3$ use two of the three $T_2$ hosts. \medskip

Returning to the general framework, we apply the rounding routine with loss target $\eta=\nicefrac{\eps}{4}$ to the retained levels corresponding to each choice of $1\le t \le r-1$.
For each  $t$, $I$ is partitioned into $H_t$, $E_{\mathrm{grp}}$, and the containment groups $\Sclass_{j,u}$.
As noted above, the first two sets are packed separately, while the objects in the containment groups correspond bijectively to the labeled rounded supply $\widehat S$. \medskip

We next introduce a hierarchy of squares for packing $\widehat S$ in a structured way.

\section{Structured Packing}
\label{sec:transfer}

Fix a separator residue $t\in \{1, \ldots, r-1\}$ and the corresponding level-separated partition of $I\setminus H_t$ provided by Lemma~\ref{lem:preprocessing} given by $(S_0, \ldots, S_J)$ and $s_0, \ldots, s_J\in(0,1]$.
Let $\widehat S$ denote the rounded supply.
Throughout this section, we use disjointness to mean interior-disjointness.
Full proofs and geometric details are given in Appendix~\ref{app:structured-packing-details}.

We construct a packing of $\widehat S$ as a rooted forest whose roots are square bins of side $s_0=1$.
Starting from level $j = 0$, we pack copies from $\widehat S_j$ inside squares of side $s_j$, and if $j<J$, we reserve squares of side $s_{j+1}$ within the squares of side $s_j$, disjoint from one another and from all already packed copies.
We then repeat this process at level $j+1$ within each reserved square.
In particular, a square may contain no copies from $\widehat S_j$ and serve only to accommodate reserved squares of side $s_{j+1}$.
At the last level, $J$, no smaller squares are reserved.
Together, the number of unit bins and the local placements must accommodate the rounded supply $\widehat S$.
This hierarchical organization, in turn, lets us assemble independently computed local packings at each level, as in the nested sub-bin organization of~\cite{MPSSW14}.

To bound the number of bins needed to pack $\widehat S$ as such a rooted forest, we first show that we can convert an unrestricted packing of $\widehat S$ into this form using only a few additional bins by repacking copies into the new bins.
%

\begin{definition}[Structured packing]
  \label{def:structured-packing}
  A \emph{structured packing} consists of a packing and a rooted forest whose nodes are axis-aligned square \emph{regions}, with the unit bins as roots.
  A region at depth $j$ is called a \emph{level-$j$ region} and has side length $s_j$.
  Each packed level-$j$ object is assigned as a \emph{current} object to one containing level-$j$ region.
  Each level-$j$ region contains level-$j+1$ regions as its children, which are pairwise disjoint and disjoint from the level-$j$ current objects.
  Level-$J$ regions have no children.
\end{definition}

In a structured packing, non-root regions are called \emph{sub-bins}, and objects assigned to descendant regions are \emph{deeper} objects.
A region is \emph{occupied} if its subtree contains an assigned object.
Child regions need not tile their parent or align with a grid.
The grids used below serve only to construct a comparison packing for the analysis.

Now, let $\Bchild:=\Mgrid^2=\nicefrac{1}{\gamma^2}$.
For $j<J$, Lemma~\ref{lem:preprocessing} gives $\nicefrac{s_j}{s_{j+1}}\in\{1,\ldots,\Mgrid\}$, so area bounds the number of children of a level-$j$ region by $\left(\nicefrac{s_j}{s_{j+1}}\right)^2\le\Bchild$.
By~\eqref{eq:K}, each region contains at most $\Kloc$ current objects.
Fix a packing of $\widehat S$ into $m$ bins as an analysis witness.
Each $P\in\widehat S_j$ is a copy of some $\widehat P_{j,u}\in S_j$, so it satisfies the diameter bounds for $S_j$ and, when $j<J$, $s_{j+1}< \gridrate(\nicefrac{\lambda\eps}{64}) \cdot b(P)$ by~\eqref{eq:child-resolution}.
If $J=0$, we keep the packing and set $\Zexc=\varnothing$.
If $J\ge1$, then we proceed with three successive deletion steps to enforce the structured packing properties.
For this comparison construction, we use the grids $\mathcal G(s_j,\tau)$, $1\le j\le J$, with one common phase $\tau\in[0, s_1)^2$ for every bin of the fixed packing.
Integer ratios between consecutive $s_j$ ensure that the grids are nested.
For $1\le j\le J$, we intersect each bin with the closures of cells of $\mathcal G(s_j,\tau)$ whose interiors intersect the bin's interior, thereby refining each bin level by level.
The bins and these intersections induce a rooted forest, with the level-$j$ intersections at depth $j$ and parents determined by containment between consecutive levels.
Some boundary regions are clipped and need not be squares, so this intermediate forest does not yet define a structured packing. \medskip

A level-$j$ copy, for $j\ge1$, is \emph{cut by its own grid} if its interior crosses a line of $\mathcal G(s_j,\tau)$.
An uncut copy lies in the closure of a unique cell at its own level, even when its boundary touches a grid line.

Following Arora’s shifted dissection averaging strategy~\cite{Aro98}, we choose $\tau$ uniformly from $[0, s_1)^2$ for the analysis.
For $1\le j\le J$, its residue modulo $s_j$ is uniform on $[0, s_j)^2$, so each $P\in\widehat S_j$ is cut with probability at most $\nicefrac{2D(P)}{s_j}\le2\alpha$.
Since the copies at lower levels have total area at most $m$, the bulk inequality~(\ref{eq:lambda-grid}) gives the following bound.

\begin{lemma}
  \label{lem:globalshift}
  There is a phase $\tau^\star\in[0,s_1)^2$ such that the multiset $\Zcut$ of lower-level copies of $\widehat S$ whose interiors cross their own level grid, which we delete in the first deletion, satisfies
  \begin{equation}
  \label{eq:cut-bound} \sum_{P\in\Zcut}D(P)^2 \le\frac{2\alpha}{\lambda}m.
  \end{equation}
\end{lemma}

With $\tau=\tau^\star$ fixed, we remove $\Zcut$ as the first deletion given by the lemma and assign each surviving lower-level copy to its containing region at its own level. \medskip

For the second deletion, we delete every surviving lower-level copy whose level-$1$ ancestor region is clipped by a bin boundary.
Let $\Zfringe$ be the multiset of copies removed in this step.
We then prune unoccupied regions, including empty roots.
In each unit bin, all clipped level-$1$ regions lie in the union of four boundary strips of width at most $s_1$, whose total area is at most $4s_1$.
We have the following bound for $\Zfringe$.

\begin{lemma}
  \label{lem:square-hierarchy}
  The multiset $\Zfringe$ of lower-level copies of $\widehat S$ deleted in the second deletion, namely those whose level-$1$ ancestor regions are clipped by bin boundaries, satisfies
  \begin{equation}
  \label{eq:fringe-bound} \sum_{P\in\Zfringe}D(P)^2 \le \frac{4s_1}{\lambda}m \le \frac{4\alpha^3}{\lambda}m.
  \end{equation}
\end{lemma}

After deletion and pruning, every retained region is a square with at most $\Bchild$ children. \medskip

\begin{figure}[t]
  \centering
  \begin{tikzpicture}[x=1.1cm,y=1.1cm,font=\small,line cap=round,line join=round,scale=0.8]

    \fill[problemcell] (2.20,2.05) rectangle (3.20,3.05);
    \draw[draw=black!68,line width=0.9pt] (1.20,1.05) rectangle (4.20,4.05);
    \foreach \x in {2.20,3.20} \draw[figfine] (\x,1.05)--(\x,4.05);
    \foreach \y in {2.05,3.05} \draw[figfine] (1.20,\y)--(4.20,\y);
    \draw[draw=aptasred,fill=white,pattern=north east lines,
        pattern color=aptasredlight,line width=0.85pt,dashed] (2.20,2.05) rectangle (3.20,3.05);
    \node[font=\footnotesize,text=black,anchor=center,color=aptasred!80!black] at (3.7,2.50) {$C \setminus P$};
    \pic[currentpoly,xscale=3.4,yscale=2.6] at (2.30,3.00) {nonconvexPolygon};
    \node[font=\large\bfseries] at (2.10,3.00) {$P$};
    \node[font=\Large,text=black] at (2.91,2.34) {$\blacktriangle$};
    \node[font=\small,text=black] at (2.5,2.25) {$\blacksquare$};

  \end{tikzpicture}
  \caption{
    In the grid-based construction, deeper copies may avoid the current copy $P$ even though $\operatorname{int}(C)\cap\operatorname{int}(P)\ne\varnothing$ for their child square $C$.
  }
  \label{fig:problematic}
\end{figure}
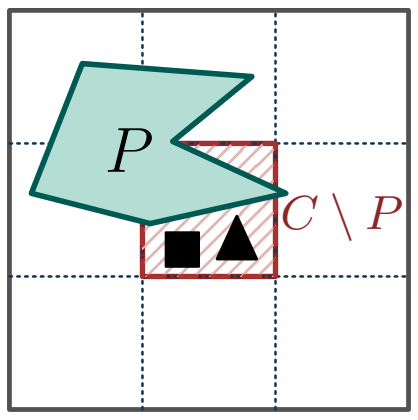

Unfortunately, from what we have shown so far, we still must deal with the following issue caused by grids determining child squares of each retained region:
A child square and a current object in the same region may overlap in their interiors, even though all deeper copies inside that region remain disjoint from this current object (see Figure~\ref{fig:problematic}), violating Definition~\ref{def:structured-packing}.
Indeed, we need its child squares, not just the deeper copies inside it, to be disjoint from current objects.
Related hierarchical grid decompositions appear in~\cite{MPSSW14}, with related constructions in~\cite{CDM23,ABGKMW24}.

To fix this problem, we process the forest top-down.
In a level-$j$ region $R$, let $X_R \subseteq \widehat S_j$ be its assigned current objects.
A level-$(j+1)$ child region $C$ is \emph{problematic} if it contains a still-present deeper object and $\operatorname{int}(C)\cap\operatorname{int}(P)\ne\varnothing$ for some $P\in X_R$.
When this happens, we choose one intersected object $P$, charge $C\setminus P$ to $P$, then delete every deeper object inside regions in the subtree rooted at $C$, delete the subtree, and stop processing it.
Since processing stops below each such child $C$, no two subtrees deleted by this rule share a region, and thus each deleted copy is charged once.

The deleted copies in a child region $C$ that create a charge on an object $P$ have total area at most $\area(C\setminus P)$, since they are interior-disjoint from $P$.
Moreover, each problematic child region deleted because of $P$ is the closure of a cell of $\mathcal G(s_{j+1},\tau)$ whose interior meets $\operatorname{int}(P)$.
Hence, by~\eqref{eq:child-resolution} and uniform rasterizability, the total charge they incur is at most $\area(V_{s_{j+1}}(P)\setminus P)\le\nicefrac{\lambda\eps}{64}\cdot \area(P)$.
The distinct current copies receiving charges have total area at most $m$, so the copies deleted in this step have total area at most $\nicefrac{\lambda\eps}{64}\cdot m$.
The bulk inequality~\eqref{eq:lambda-grid} then yields the following bound.

\begin{lemma}
  \label{lem:problematic}
  The multiset $\Zprob$ of lower-level copies of $\widehat S$ deleted when removing problematic children in the third deletion satisfies
  \begin{equation}
  \label{eq:problem-bound} \sum_{P\in\Zprob}D(P)^2 \le\frac{\eps}{64}m.
  \end{equation}
\end{lemma}

Pruning unoccupied regions therefore leaves a structured packing of surviving rounded copies using at most $m$ initial bins, with at most $\Bchild$ children per region and, by~\eqref{eq:K}, at most $\Kloc$ current objects per region.
Also, every surviving copy retains its initial position, and since the three successive deletion multisets are disjoint, their union $\Zexc:=\Zcut\cup\Zfringe\cup\Zprob$ satisfies
\begin{equation}
    \label{eq:transfer-exception}
    \sum_{P\in\Zexc}D(P)^2 \le
    \left(\frac{2\alpha}{\lambda}+\frac{4\alpha^3}{\lambda}+\frac{\eps}{64}\right) \cdot m.
\end{equation}
None of the three deletion steps affects level $0$, so every deleted copy $P \in \Zexc$ belongs to a level $j\ge1$ and thus satisfies $D(P)\le\alpha\cdot s_j\le\alpha<\nicefrac{1}{2}$. \medskip

Our comparison packing must still pack all copies in $\Zexc$.
We use the following lemma to bound the number of additional structured bins needed.
The proof follows from a straightforward application of NFDH.

\begin{lemma}
  \label{lem:structured-completion}
  Under the parameter choices of Section~\ref{sec:abstract-geometry}, every submultiset $Z$ of lower-level copies of $\widehat S$ admits a structured packing in at most
  \[
    \frac{2}{1-\alpha}\sum_{P\in Z}D(P)^2+\frac{1}{1-\alpha}
  \]
  unit bins where each region contains at most $\Kloc$ current objects and has at most $\Bchild$ children.
\end{lemma}

To complete the conversion of the starting packing of $\widehat S$ into a structured packing, we apply Lemma~\ref{lem:structured-completion} to $\Zexc$.
It then follows that the entire rounded supply $\widehat S$ has a structured packing in at most $m+\frac{2}{1-\alpha}\sum_{P\in\Zexc}D(P)^2+\frac{1}{1-\alpha}$ bins.
We stress that the initial rounded packing, grid phase, multiset of deleted items, and comparison structured packing are all analysis-only witnesses.

\section{Algorithm and approximation guarantee}
\label{sec:local-configurations}
\label{sec:approx}

For fixed $0<\eps<1$ and a separator residue, the preceding section bounded the bins needed for a structured packing of its corresponding rounded supply $\widehat S$.
We now turn to computing an optimal such packing.
We then establish the approximation guarantee using the deletion bound~\eqref{eq:transfer-exception} and Lemma~\ref{lem:structured-completion}.
Full proofs and further details are given in Appendix~\ref{app:structured-packing-details}.

At level $j$, let $\mathcal U_j$ index the containment groups $\Sclass_{j,u}$.
Each group contributes $n_{j,u}:=|\Sclass_{j,u}|$ labeled copies of its chosen maximum $\widehat P_{j,u}$ to $\widehat S_j$.
Also, $\widehat S$ contains at most $n$ labeled copies. \medskip

We first determine what can be packed in single square regions.
Because every level-$j$ region is a translate of $R_j=[0,s_j]^2$, we compute local packings in $R_j$ for each level.
At level $j$, we use the algorithm of Lemma~\ref{lem:local-solver} to query whether prescribed numbers of current rounded copies of each $\widehat P_{j,u}$ can be packed in $R_j$, together with a prescribed number of squares of side $s_{j+1}$ for $j<J$.
When $j<J$, these squares reserve space for child square regions and are not rounded copies, so they are not counted toward $n_{j,u}$.
Our local queries allow every placement of reserved child squares permitted by Definition~\ref{def:structured-packing}, without requiring grid alignment.

To solve the global problem, we build a single integer program with a separate set of variables for each level.
We adapt the classical configuration formulation of bin packing~\cite{GG61}: each level $j$ variable counts regions using a configuration of current copies from $\widehat S_j$.
Accordingly, each configuration records only the numbers of current copies in one region of a given level, rather than the entire contents of a unit bin.
Furthermore, for every $j<J$, we add capacity constraints to ensure enough reserved child squares in level-$j$ regions to accommodate all level-$(j+1)$ regions.

For each level $j$, a \emph{local configuration} is a vector $\mathbf{c}=(c_u)_{u\in\mathcal U_j}$ specifying how many rounded copies from each containment group to pack as current objects in one level-$j$ region, with $c_u\in\mathbb Z_{\ge0}$.
Here $c_u$ specifies the multiplicity for $\Sclass_{j,u}$.
The occupancy bound~\eqref{eq:K} lets us consider only vectors with $|\mathbf{c}|_1=\sum_{u\in\mathcal U_j}c_u\le\Kloc$.
Moreover, different containment groups have separate coordinates in $\mathbf{c}$, even when their chosen maxima have identical geometry.
Let $\mathcal X_j$ be the set of configurations whose current copies fit in $R_j$.
Since $|\mathcal U_j|\le\Rrep$, stars and bars gives $|\mathcal X_j|\le\Hpat:=\binom{\Rrep+\Kloc}{\Kloc}$.
We keep nonmaximal configurations because adding current copies can reduce the number of child squares that fit together.
For $\mathbf{c}\in\mathcal X_j$, the \emph{local capacity} $\kappa_j(\mathbf{c})$ is the maximum number of squares of side $s_{j+1}$ packable jointly with its current copies in $R_j$ for $j<J$, and $\kappa_J(\mathbf{c})=0$.
In particular, for $j<J$, the empty configuration $\mathbf{0}\in\mathcal X_j$ has $\kappa_j(\mathbf{0})=\left(\nicefrac{s_j}{s_{j+1}}\right)^2$, allowing regions to carry deeper copies through levels with empty supply.

Using Lemma~\ref{lem:local-solver}, we now compute every $\mathcal X_j$.
For $\mathbf{c}\in\mathcal X_j$ and $j<J$, feasibility is downward closed in the number of child squares, so we can binary search over $0, \dots, \Bchild$ to determine $\kappa_j(\mathbf{c})$.
Each query contains at most $\Kloc+\Bchild$ movable objects, so we can compute $\mathcal X_j$, its capacities, and one rational maximizing realization per configuration in polynomial time.
Moreover, because we can translate entire packings within a level-$(j+1)$ region into any reserved square of side $s_{j+1}$, only total child capacity matters.
The following lemma makes these claims precise.

\begin{lemma}
  \label{lem:composition}
  In each occupied level-$j$ region of a structured packing of $\widehat S$, the current copies define a local configuration $\mathbf{c}=(c_u)_{u\in\mathcal U_j}\in\mathcal X_j$, and the region has at most $\kappa_j(\mathbf{c})$ occupied children.

  Similarly, suppose that each depth-$j$ node of a finite rooted forest of depth at most $J$ is assigned a configuration $\mathbf{c}\in\mathcal X_j$ and has at most $\kappa_j(\mathbf{c})$ children.
  Then the rounded copies specified by these configurations admit a structured packing in at most one bin per root, thereby realizing the forest after pruning unoccupied subtrees.
  Each retained depth-$j$ node assigned $\mathbf{c}$ corresponds to a region with $c_u$ current copies of $\widehat P_{j,u}$ for every $u\in\mathcal U_j$.
\end{lemma}

By Lemma~\ref{lem:composition}, we can optimize the bin count by choosing how often each local configuration is used.
Let $x_{j,\mathbf{c}}$ be a variable counting level-$j$ regions using configuration $\mathbf{c}$.
We require a total of exactly $n_{j,u}$ copies of $\widehat P_{j,u}$ for every $j$ and $u\in\mathcal U_j$.
Then, for every $j<J$, we ensure that the total number of level-$(j+1)$ regions is at most the total child capacity of the level-$j$ regions, so the next-level regions fit into the reserved child squares.
Finally, we minimize the number of level-$0$ regions, that is, the unit bins.

\begin{lemma}
  \label{lem:structured-selection}
  For a separator residue, we can compute an optimal structured packing of $\widehat S$ in polynomial time.
  Moreover, we can pack the objects of $I$ replaced by copies in $\widehat S$ using at most as many bins as any such optimal structured solution.
\end{lemma}

In any structured packing of $\widehat S$, removing unoccupied regions leaves at most $n$ regions per level, so each $x_{j,\mathbf{c}}$ can be restricted to $\{0,\ldots,n\}$.
Each constraint of the integer program involves variables from one level $j$ or from two consecutive levels $j$ and $j+1$.
Each level contributes at most $\Hpat$ variables, so the graph joining variables that share a constraint has a path decomposition of width at most $2\Hpat-1$.
For each residue, we solve the integer program using an optimization variant of Jansen and Kratsch's bounded-domain feasibility dynamic program~\cite{JK15}. \medskip

Finally, to prove the approximation guarantee, fix an input $I$ drawn from $\calP$ and a residue $t$ supplied by Lemma~\ref{lem:medium}.
Representative roundability ensures that the full rounded supply $\widehat S$ packs into $\OPT(\widehat S) \le \OPT(I)$ bins.
By Lemma~\ref{lem:structured-selection}, we can pack the corresponding objects of $I$ using no more bins than the optimal structured packing.
 Adding the separately computed packings of $H_t$ and $E_{\mathrm{grp}}$, and selecting the best residue, gives a final bin count $A_\eps(I)$ at most
\begin{equation}
\label{eq:final-abstract}
\left(1+\frac{4}{\lambda(r-1)}+\frac{\eps}{4}
+\frac{2}{1-\alpha}\left(\frac{2\alpha}{\lambda}+\frac{4\alpha^3}{\lambda}+\frac{\eps}{64}\right)\right)\cdot\OPT(I)
+2+\delta_{\mathrm{grp}}+\frac{1}{1-\alpha}.
\end{equation}
Here $\delta_{\eps,\calP}:=\left\lceil 2+\delta_{\mathrm{grp}}+\frac{1}{1-\alpha}\right\rceil$.
The estimates in Appendix~\ref{app:parameter-budget} show that the coefficient of $\OPT(I)$ in~\eqref{eq:final-abstract} is at most $1+\eps$.
Together with the rational reconstruction and running-time analysis in Appendix~\ref{app:selection-runtime}, we therefore complete the proof of Theorem~\ref{thm:abstract-main}.

\section{Conclusion}
\label{sec:conclusion}

We established an APTAS for translation-only geometric bin packing of any fixed class of rational polygonal regions satisfying uniform rasterizability and uniform representative roundability.
The framework also handles non-convex objects and holes, with no fixed bound on boundary complexity, and returns exact rational packings without bin augmentation.

In the same spirit, the framework extends to fixed dimension $d\ge2$ for compact rational polyhedral regions with nonempty connected interiors, represented by simplicial decompositions.
Hence grids use hypercubes, area becomes $d$-dimensional volume, and $b(P)$ is the maximum coordinate span, while $D(P)$ remains the Euclidean diameter.
Accordingly, we state the following result, a direct extension of Theorem~\ref{thm:abstract-main}.

\begin{theorem}
  \label{thm:higher-dimensional}
  For every fixed integer $d\ge2$ and fixed class $\calP$ of these regions satisfying the higher dimensional versions of uniform rasterizability and uniform representative roundability, and for every instance $I\subseteq\calP$ of binary encoding length $L$ and $0<\eps<1$, we can compute a translation-only packing into at most
  \[
    (1+\eps)\OPT(I)+\delta_{\eps,\calP,d}
  \]
  unit hypercubes, with exact rational translations of polynomial encoding length, in time $L^{g_{\calP,d}(1/\eps)}$.
  Here $g_{\calP,d}$ is computable, and $\delta_{\eps,\calP,d}$ is independent of the instance, $n$, and $L$.
\end{theorem}

The extension follows immediately by using volume, hypercubical grids, dyadic bounding hypercubes, and simplicial decompositions in the same arguments, with dimension-dependent constants.

\paragraph{Use of generative AI.}

The authors developed the main ideas and proof strategies.
We used OpenAI Codex broadly to explore these ideas and fill in technical proof details, such as computing constants and parameter bounds.
We also used it to propose tests and to draft and edit the exposition.
The authors verified the correctness and originality of all content, including references, and take full responsibility for the final paper.

\clearpage
\appendix
\section{Grid geometry and bulk}
\label{app:grid-geometry}

Below, we prove the relationship between uniform rasterizability and uniform relative outer Minkowski growth, and derive the bulk inequality used throughout the framework.

\subsection{Uniform rasterizability and uniform relative outer Minkowski growth}
\label{app:minkowski-proofs}

\label{def:parallel-sets}
For a nonempty compact set $A\subseteq\mathbb R^n$ and $\delta>0$, write its closed $\delta$-neighborhood as
  \[
    A_\delta:=\{x\in\mathbb R^n \mid \operatorname{dist}(x,A)\le\delta\}=A+\callB(0,\delta),
  \]
where $+$ denotes the Minkowski sum and $\callB(0,\delta)$ is the closed Euclidean ball of radius $\delta$~\cite[Chap.~3]{Fal03}.
For every fixed polygonal region $P$, the outer increment $\area(P_\delta\setminus P)$ tends to zero as $\delta\to0^+$.
We instead require a relative neighborhood radius that gives an excess-area bound uniformly over $\calP$.

\begin{definition}[Uniform relative outer Minkowski growth]
  \label{def:outer-parallel}
  $\calP$ has \emph{uniform relative outer Minkowski growth} if there is a polynomial-time computable function $\Sigma\colon (0,1)\cap\mathbb Q\to(0,1]\cap\mathbb Q$ such that the following holds.
  For every rational $0<\eta<1$, every $P\in\calP$, and every radius $0<\delta\le\Sigma(\eta)\cdot b(P)$, $P_\delta$ satisfies
  \begin{equation}
  \label{eq:outer-parallel} \area\bigl(P_\delta\setminus P\bigr) \le\eta\cdot \area(P).
  \end{equation}
\end{definition}

We call $\Sigma$ the \emph{outer-resolution function}.
Uniform relative outer Minkowski growth is formulated using continuous Euclidean neighborhoods, whereas the transfer analysis uses square-grid cells.
The inclusions
\[
  \callB(0,\delta)\subseteq[-\delta,\delta]^2
  \subseteq\callB(0,\sqrt2\cdot \delta)
\]
show that Euclidean parallel sets and square neighborhoods differ only by constant factors in their radii.
The grid condition in Definition~\ref{def:gridregular} uses a cell side length in place of the continuous neighborhood radius and requires the excess-area guarantee for every phase.

\begin{proposition}
  \label{prop:grid-parallel}
  Uniform rasterizability and uniform relative outer Minkowski growth are equivalent up to constant rescaling.
  More precisely:
  \begin{enumerate}[label=(\roman*)]
    \item Definition~\ref{def:outer-parallel} implies Definition~\ref{def:gridregular} with the rational grid-resolution function $\gridrate(\eta)=\nicefrac12\Sigma(\eta)$.
    \item Definition~\ref{def:gridregular} implies Definition~\ref{def:outer-parallel} with
          \begin{equation}
          \label{eq:delta-from-grid} \Sigma(\eta)=\nicefrac{1}{8}\gridrate(\nicefrac{\eta}{64}).
          \end{equation}
  \end{enumerate}
\end{proposition}

\begin{proof}[Proof of Proposition~\ref{prop:grid-parallel}]
  For (i), every point of a selected cell lies at a distance of at most $\sqrt{2}\cdot h$ from an interior point of $P$ certifying the cell's selection.
  Hence, $V_{h,\tau}(P)\setminus P\subseteq P_{\sqrt2h}\setminus P$.
  Thus $h\le\nicefrac{\Sigma(\eta)b(P)}{2}$ gives $\sqrt2\cdot h\le\Sigma(\eta)\cdot b(P)$, so \eqref{eq:outer-parallel} implies \eqref{eq:outer-abstract}.

  For (ii), fix $\delta>0$ with $8\delta\le\gridrate(\nicefrac{\eta}{64})\cdot b(P)$ and put $H:=8\delta$.
  Consider the $8\times8$ phase family $\mathcal T_\delta:=\{(i\delta,j\delta) \mid i,j\in\{0,\ldots,7\}\}$ for grids with cell side length $H$.
  If $x\in P_\delta$, choose $y\in P$ with $\lVert x-y\rVert_2\le\delta$, hence also $\lVert x-y\rVert_\infty\le\delta$.
  Since every polygonal region is the closure of its interior, choose $y'\in\operatorname{int}(P)$ with $\lVert y-y'\rVert_\infty<\delta$.
  For coordinate $c = 1$ or $2$, the closed interval between $x_c$ and $y'_c$ has length strictly less than $2\delta=\nicefrac{H}{4}$.
  Among the eight line families $i\delta+H\mathbb Z$, $i\in\{0,\ldots,7\}$, this interval can meet grid lines from at most three distinct families.
  Thus each coordinate has an offset with no grid line in the interval, and choosing the two offsets independently places $x$ and $y'$ in one common open $H\times H$ cell.
  Since that cell contains $y'\in\operatorname{int}(P)$, its closure contributes to $V_{H,\tau}(P)$ and contains $x$.
  Hence
  \[
    P_\delta\setminus P \subseteq\bigcup_{\tau\in\mathcal T_\delta}\bigl(V_{H,\tau}(P)\setminus P\bigr).
  \]
  The shifted-grid bound at accuracy $\nicefrac{\eta}{64}$ holds for each of the $8^2=64$ phases, because $H\le\gridrate(\nicefrac{\eta}{64})\cdot b(P)$.
  Therefore, subadditivity and the above containment give
  \[
    \area(P_\delta\setminus P) \le\sum_{\tau\in\mathcal T_\delta}\area(V_{H,\tau}(P)\setminus P) \le 64\cdot \nicefrac{\eta}{64}\area(P)=\eta\cdot \area(P).
  \]
  This proves \eqref{eq:outer-parallel} whenever $\delta\le\Sigma(\eta)\cdot b(P)$ for $\Sigma$ defined in \eqref{eq:delta-from-grid}.
\end{proof}

\subsection{Bulk from rasterizability}
\label{app:grid-bulk}

\begin{proof}[Proof of Proposition~\ref{prop:grid-bulk}]
  Put $c:=\min\{\gridrate(\nicefrac{1}{2}),\nicefrac{1}{2}\}$ and use an arbitrary phase with cell side length $h:=c \cdot b(P)$.
  After interchanging the coordinate axes if necessary, assume that the horizontal projection of $P$ has length $b(P)$.
  The coordinate projection is continuous, so connectedness of $P$ implies that its image is one interval of length $b(P)$ rather than a union of separated intervals.
  Polygonal holes do not affect this connectedness argument.
  The projection of $\operatorname{int}(P)$ is dense in this interval.
  Every one of at least $\lfloor\nicefrac{b(P)}h\rfloor$ grid-column interiors that meet it therefore contains the first-coordinate projection of an interior point of $P$.
  Since $\operatorname{int}(P)$ is open, that point can be chosen away from horizontal grid lines.
  Thus the grid-cell cover has a selected cell in each of those columns.
  Taking one touched cell from each such column gives cells with disjoint interiors.
  Since $c\le\nicefrac{1}{2}$,
  \[
    \area(V_{h,\tau}(P))\ge \left\lfloor\frac{b(P)}h\right\rfloor h^2 \ge\frac{b(P)}{2h}h^2 =\frac{c \cdot b(P)^2}{2}.
  \]
  The shifted-grid guarantee at $\eta=\nicefrac{1}{2}$ gives $\area(V_{h,\tau}(P))\le\nicefrac{3}{2} \area(P)$.
  Hence $\area(P)\ge\frac{c \cdot b(P)^2}{3}$.
  Finally $D(P)^2\le2\cdot b(P)^2$, so
  \[
    \area(P)\ge\frac{c \cdot b(P)^2}{3}\ge\frac{c \cdot D(P)^2}{6} =\lambda\cdot D(P)^2,
  \]
  which proves \eqref{eq:lambda-grid}.
\end{proof}

\section{Preprocessing}
\label{app:preprocessing-rounding}

In this appendix, we present the construction of a level-separated partition, the representative rounding procedure, and the approximation factor analysis of our algorithm.

\subsection{Construction of a level-separated partition}
\label{app:scale-construction}

We now determine the appropriate values of $r, \alpha, \beta, \gamma$ to prove Lemma~\ref{lem:preprocessing}.
The construction must simultaneously:
bound the total area of the separator set, determine appropriate size of square regions for lower-level objects, and ensure a grid scale fine enough in each level to apply uniform rasterizability.

Let $\lambda$ be the grid-bulk constant from Proposition~\ref{prop:grid-bulk}.
We use rounding loss $\nicefrac{\eps}{4}$ and grid-area error $\nicefrac{\lambda\eps}{64}$.
We use $r=\lceil \frac{32}{\lambda\cdot\eps} \rceil+2$ and set $\beta:=\alpha^{2r-1}$, $\gamma:=\alpha\cdot \beta = \alpha^{2r}$, and choose a reciprocal power of two $\alpha=2^{-K}$ sufficiently small so that
\begin{equation}
\label{eq:alpha-choice}
\alpha\le\frac{\lambda\cdot\eps}{128} \qquad
\text{and} \qquad
\alpha\le\frac{\gridrate(\nicefrac{\lambda\eps}{64})}{\sqrt2}.
\end{equation}

These bounds are relevant for the proof of the existence of a structured packing of the rounded supply. 
The first bound on $\alpha$ controls the own-grid cut and boundary-fringe contributions.
The second bound on $\alpha$, together with $D(P)\le\sqrt2\cdot b(P)$, gives the child-grid resolution needed to bound the problematic-child deletion charge by grid-cell-cover excess.
Moreover, $\beta$ provides a diameter lower bound to control the maximum occupancy of the square containers at each level.
Furthermore, at each nonterminal lower level, $\gamma=\alpha\cdot \beta$ gives $s_{j+1}=\gamma\cdot s_j<\alpha\cdot D(P)$ for $P\in S_j$, and the dyadic scales give integer subdivision ratios.

Recall the sets $G_i$, separator sets $H_t$, and set $L_{>1}$ from Section~\ref{sec:abstract-geometry}.
For $t\in\{1,\ldots,r-1\}$, $H_t$ is the union of every $r$-th set $G_i$ beginning with $G_t$, as shown in Figure~\ref{fig:scale-separation}.
Objects in $L_{>1}$ belong to level $0$, whose regions are unit bins and have no upper diameter requirement of the form $\alpha\cdot s_0$.
After selecting $H_t$ for separate packing, the retained levels are
\begin{equation}
S_0 :=L_{>1}\cup\bigcup_{i=0}^{t-1}G_i, \qquad
\text{and} \qquad
S_j :=\bigcup_{i=t+(j-1)r+1}^{t+jr-1}G_i\quad(j\ge1).
\end{equation}
The retained levels $S_0, S_1, \dots$ partition the input objects not in $H_t$.
We take $J$ to be the largest index with $S_J\ne\varnothing$ if such an index exists, and $J=0$ otherwise.
We retain empty intermediate levels to simplify exposition.
The associated positive level parameters are $s_0:=1$ and $s_j:=\alpha^{2(t+(j-1)r)+1}$ for $j\ge1$.
The following calculation verifies that each retained level fits the diameter window of Definition~\ref{def:separated}.
For $j\ge1$, we have that
\begin{equation}
\label{eq:scale-window} \beta\cdot s_j<D(P)\le \alpha\cdot s_j\quad(P\in S_j), \quad \text{and} \quad s_{j+1}=\gamma\cdot s_j.
\end{equation}

Indeed, $s_{j+1}=\alpha^{2r}\cdot s_j=\gamma\cdot s_j$ and $\beta s_j=\alpha^{2(t+jr)}$.
For each $j\ge1$, the indices of the sets $G_i$ forming $S_j$ range from $t+(j-1)r+1$ to $t+jr-1$, giving the two diameter bounds in \eqref{eq:scale-window}.
The next set $G_{t+jr}$ is included in $H_t$ and separates this retained level from the next one.
This gap of one set $G_i$ is what permits both the lower diameter bound at level $j$ and the smaller cell side length required at level $j+1$.
For $j=0$, the definitions give a stronger bound on the diameter of the objects in the top level: For each $P\in S_0$, we have 
\begin{equation}
\label{eq:top-window} D(P)>\alpha^{2t}>\beta\cdot s_0.
\end{equation}
Thus $(S_0,\ldots,S_J)$, together with $s_0,\ldots,s_J$, is an $(\alpha,\beta,\gamma)$-level-separated partition of $I\setminus H_t$.

For $J\ge1$, the root ratio is $s_0/s_1=\alpha^{-(2t+1)}<\alpha^{-2r}=\Mgrid$, and all later consecutive ratios equal $\Mgrid$.
They are integers because $\alpha$ is a reciprocal power of two.
Also $s_1\le\alpha^3$ and $\gamma \le \nicefrac{1}{2}$.
For a current object at a nonterminal level $j\ge1$, $s_{j+1}=\gamma\cdot s_j=\alpha\cdot \beta\cdot s_j<\alpha\cdot D(P)$.
At the root, \eqref{eq:top-window} gives $s_1=\alpha^{2t+1}<\alpha\cdot D(P)$.
Together with \eqref{eq:bdiam} and \eqref{eq:alpha-choice}, this proves \eqref{eq:child-resolution}.

Every positive-area input polygon has two boundary vertices differing in some coordinate.
A nonzero difference of rational input coordinates has magnitude at least $2^{-O(L)}$, so $D(P)\ge2^{-O(L)}$ for every input object.
If $J\ge1$, an object in the deepest retained level satisfies
\[
  2^{-O(L)}\le D(P)\le\alpha\cdot s_J
  =\alpha\cdot s_1\cdot \gamma^{J-1}\le2^{-(J-1)},
\]
since $\gamma=\alpha^{2r}\le\nicefrac12$ and $\alpha\cdot s_1\le1$.
Thus $J=O(L)$, including empty intermediate levels.
The convention $J=0$ handles the absence of lower levels.

The scales $s_0=1$ and $s_j=\alpha^{2(t+(j-1)r)+1}$, for $j\ge1$, have $F(1/\eps)\cdot O(L+1)$ bits, where $F$ depends only on $\calP$.
All input labels are assigned by diameter comparisons to exactly one retained level or to $H_t$.
The $r-1$ separator residues depend only on $\calP$ and $\eps$.
Empty input and the absence of lower-level objects use $J=0$.
These facts prove Lemma~\ref{lem:preprocessing}.

Next, we prove the bounds in Lemma~\ref{lem:medium} on the separator area and the number of unit bins needed to pack its objects.

\subsection{Separator residue}
\label{app:separator}

The following lemma bounds the number of unit bins needed to pack a set of squares. 
This standard bound follows from the analysis of Coffman, Garey, Johnson, and Tarjan~\cite{CGJT80}.

\begin{lemma}
  \label{lem:nfdh}
  Let $\mathcal S$ be a multiset of axis-aligned squares of total area $A$ and maximum side $\eta<1$.
  Next-Fit Decreasing Height (NFDH) packs them into at most $\frac{2A+\eta}{1-\eta}+1$ unit-square bins.
\end{lemma}

In particular, Lemma~\ref{lem:nfdh} implies that if $\eta\le\nicefrac{1}{2}$ then at most $4A+2$ bins suffice.
We use this result to show that the objects in a separator set can be packed into few unit square bins. 

\begin{proof}[Proof of Lemma~\ref{lem:medium}]
  Since the packed interiors are disjoint and the bins have total area $\OPT(I)$, we have $\area(I)\le \OPT(I)$.
  The $r-1$ separator sets are disjoint, so at least one residue satisfies \eqref{eq:cheap-residue}.

  Fix any residue satisfying this bound.
  We place every $P\in H_t$ in its axis-aligned bounding square.
  By the span-diameter relation \eqref{eq:bdiam}, the bulk inequality, and the area bound on the separator set \eqref{eq:cheap-residue}, we have
  \[
    \sum_{P\in H_t}b(P)^2 \le\sum_{P\in H_t}D(P)^2 \le\frac1\lambda\sum_{P\in H_t}\area(P) =\frac{\area(H_t)}\lambda \le\frac{\OPT(I)}{\lambda\cdot(r-1)}.
  \]
  Every object in the separator set has $b(P)\le D(P)\le \alpha^2\le\nicefrac{1}{2}$.
  Thus, by Lemma~\ref{lem:nfdh}, applying NFDH to their bounding squares gives a packing in at most $\frac{4\OPT(I)}{\lambda\cdot(r-1)}+2$ bins. From that, to pack the separator set itself, it suffices to place each object in the area occupied by its bounding square.
  NFDH uses polynomially many rational comparisons and additions, so this construction runs in polynomial time and returns rational translations of polynomial encoding length.
  Finally, since $r = \left\lceil\frac{32}{\lambda\cdot\eps}\right\rceil+2$, we have 
  \[
    r-1=\left\lceil\frac{32}{\lambda\cdot\eps}\right\rceil+1 \ge\frac{32}{\lambda\cdot\eps} \text{, hence } \frac4{\lambda\cdot(r-1)}\le\frac{\eps}{8},
  \]
  proving both inequalities in \eqref{eq:medium-cost}.
\end{proof}

\subsection{Quantitative rounding from translation-containment width}
\label{app:width-rounding}

The proof first turns containment comparisons into executable local queries and uses them to compute chains.
Grouping along those chains gives chosen maxima for substitution, while assigning copies to distinct hosts in preceding segments proves preservation of packability.
We then bound the number of chosen maxima per level and explicitly pack the union of the first segments separately.

\begin{proof}[Proof of Lemma~\ref{lem:width-roundable}]
  Suppose $\calP$ is uniformly rasterizable and has translation-containment width at most $d$.
  Fix a loss target $0<\eta<1$ and an $(\alpha,\beta,\gamma)$-level-separated partition $(S_0,\ldots,S_J)$ of an input multiset $I\subseteq\calP$.
  Let $\lambda$ be the bulk constant from \eqref{eq:lambda-grid} and $\Kloc$ be the bound in~\eqref{eq:K} on the number of level-$j$ objects that fit in a square of side $s_j$.

  \paragraph{Chain partition.}
  At each nonempty level, we compute the translation-containment comparison matrix on its distinct normalized shapes.
  Dilworth's theorem states that the minimum number of chains covering a finite partial order equals its width~\cite{Dil50}.
  The standard bipartite-matching construction therefore gives a partition into at most $d$ chains.
  We order every chain so that each earlier shape contains a translate of every later shape and expand each shape into its consecutive labeled occurrences.
  This yields a partition of the labeled level multiset into at most $d$ containment chains, even when some input objects have equal normalized shapes.
  Empty levels have no chains or containment groups and contribute no objects for separate packing.

  \paragraph{Grouping and returned data.}
  Each chain contributes its first segment to the set of input objects selected for separate packing and one chosen maximum per later segment retained for rounding.
  Taking a segment size proportional to the level cardinality bounds the number of segments independently of that cardinality, while the proportionality factor controls how many input objects are packed separately.
  The choice below uses the occupancy bound $\Kloc$ to relate cardinalities to packed area.
  The factor $5$ allocates the target loss between top-level packing and the NFDH bound for lower levels.
  For each nonempty level $j$, we put $n_j:=|S_j|$ and choose
  \[
    g_j:=\max\left\{1,\left\lceil\frac{\eta\cdot n_j}{5\Kloc\cdot d}\right\rceil\right\}.
  \]
  We split every chain at that level into consecutive segments of size $g_j$, starting with its first object, with only the final segment allowed to be shorter.
  For a chain of size $n_{j,c}$, write the segments as $T_1,\ldots,T_\nu$, where $\nu=\lceil n_{j,c}/g_j\rceil$, $|T_a|=g_j$ for $a<\nu$, and $|T_\nu|=n_{j,c}-(\nu-1)g_j$.
  Let $E_{\mathrm{grp}}$ be the union of the first segments $T_1$ over all chains and levels.
  The routine selects these input objects for separate packing.
  If a chain has at most $g_j$ objects, it consists only of $T_1$ and contributes no containment group.

  We index the later segments across all chains at level $j$ by $u$.
  For the segment $T_a$ with index $u$, we set $\Sclass_{j,u}:=T_a$ and choose its first member $\widehat P_{j,u}$ as the local maximum.
  The routine returns one labeled copy of $\widehat P_{j,u}$ for every input object $P\in\Sclass_{j,u}$.
  The chain order gives
  \[
    P\preceq\widehat P_{j,u}\qquad(P\in\Sclass_{j,u}).
  \]
  This containment permits substitution of an input object into a later computed placement of a rounded copy.
  The routine also returns an explicit separate packing of $E_{\mathrm{grp}}$, constructed below.

  \paragraph{Preservation of packability.}
  Fix any packing of $I$ in $m$ bins solely as an analysis witness.
  In each chain, every segment $T_a$ with $a\ge2$ has at most $g_j$ members, while its preceding segment $T_{a-1}$ has exactly $g_j$ members.
  Thus there is an injection $\pi_a:T_a\to T_{a-1}$, obtained, for example, by assigning the $k$-th member of $T_a$ to the $k$-th member of $T_{a-1}$.
  For every containment group $\Sclass_{j,u}=T_a$, the chain order gives the second containment,
  \[
    \widehat P_{j,u}\preceq\pi_a(P)\qquad(P\in\Sclass_{j,u}),
  \]
  which is used only to prove packability of the rounded supply.
  We place the copy of $\widehat P_{j,u}$ labeled by $P$ inside the placement of $\pi_a(P)$ in the fixed witness packing.

  All host labels are distinct: each $\pi_a$ is injective, different segments use disjoint preceding segments, and different chains and levels contain disjoint sets of input labels.
  Input objects with equal normalized shapes but different labels therefore remain distinct hosts.
  We replace all placements of input objects in the witness packing simultaneously by the placements of these copies.
  The input objects used as hosts are not retained in this rounded packing.
  The rounded copies lie in distinct host placements, so their interiors remain disjoint and every copy remains in its host's bin.
  This packs the entire rounded supply into the same $m$ bins, even when final segments have fewer than $g_j$ members or some chains contribute no rounded copies.
  The routine neither computes nor receives the witness packing or its number of bins $m$.

  \paragraph{Number of chosen maxima.}
  Consider a nonempty level $j$, and let $\Ncls_j$ be its number of containment groups.
  A chain containing $n_{j,c}$ objects creates at most $\lceil\nicefrac{n_{j,c}}{g_j}\rceil$ segments.
  Writing $q_j\le d$ for the number of nonempty chains,
  \[
    \Ncls_j\le\sum_{c=1}^{q_j}\left\lceil\frac{n_{j,c}}{g_j}\right\rceil \le\sum_{c=1}^{q_j}\left(\frac{n_{j,c}}{g_j}+1\right) \le\frac{n_j}{g_j}+d.
  \]
  If $g_j=1$, then $n_j\le\nicefrac{5\Kloc d}{\eta}$.
  Otherwise $g_j\ge\nicefrac{\eta n_j}{5\Kloc d}$, so in either case $\nicefrac{n_j}{g_j}\le\nicefrac{5\Kloc d}{\eta}$.
  Thus
  \begin{equation}
  \label{eq:R} \Ncls_j\le d+\frac{n_j}{g_j} \le d+\frac{5\Kloc d}{\eta}, \qquad \Rrep:=\left\lceil d+\frac{5\Kloc d}{\eta}\right\rceil.
  \end{equation}

  An empty level has $\Ncls_j=0$, so the same bound holds there.

  \paragraph{Separate packing.}
  Let $e_j$ be the number of objects in $E_{\mathrm{grp}}$ at level $j$.
  Each chain contributes at most $g_j$ such objects, and $g_j\le \frac{\eta\cdot n_j}{5\Kloc\cdot d}+1$, so for a nonempty level
  \begin{equation}
  \label{eq:deleted-count} e_j\le d g_j\le\frac{\eta\cdot n_j}{5\Kloc}+d.
  \end{equation}
  For an empty level, $e_j=n_j=0$, so the final bound in \eqref{eq:deleted-count} still holds.
  At level $0$, the lower diameter bound and bulk imply that every object has area at least $\lambda\cdot \beta^2$.
  The $m$ witness bins have total area $m$, so $n_0\le\frac{m}{\lambda\cdot \beta^2}\le\Kloc\cdot m$.
  Hence $e_0\le\nicefrac{\eta m}{5}+d$.
  We pack each top-level object in $E_{\mathrm{grp}}$ in its own bin, using its normalized coordinates.
  Every input object fits a unit bin by the standing assumption in Section~2.

  For $j\ge1$, we charge each level-$j$ object $P\in E_{\mathrm{grp}}$ an amount $s_j^2$, which bounds its bounding-square area since $b(P)\le\alpha\cdot s_j\le s_j$.
  Every object in $S_j$ has area at least $\lambda\cdot \beta^2\cdot s_j^2$, hence $n_j\cdot s_j^2\le\frac{\area(S_j)}{\lambda\cdot \beta^2}$.
  Because the parts of the partition are disjoint and $\area(I)\le m$,
  \begin{align*}
  \sum_{j=1}^J e_j\cdot s_j^2
   & \le \sum_{j=1}^J\left(\frac{\eta\cdot n_j}{5\Kloc}+d\right)\cdot s_j^2 \\
   & \le \frac{\eta}{5} \cdot m+\frac{d}{1-\gamma^2}.
  \end{align*}
  Here we used $\lambda\cdot \beta^2\cdot \Kloc\ge1$ and $\sum_{j=1}^J s_j^2\le1/(1-\gamma^2)$.
  For $J\ge1$, we use $s_1\le1$ and geometric decay.
  For $J=0$, the sum is empty.
  Every lower-level object $P\in E_{\mathrm{grp}}$ satisfies $b(P)\le D(P)\le\alpha\cdot s_j\le\nicefrac{1}{2}$, since $\alpha=2^{-K}$ for a positive integer $K$ and $s_j\le1$.
  We apply NFDH from Lemma~\ref{lem:nfdh} to their bounding squares and place each input object inside its square, using at most
  \[
    \frac{4\eta}{5}\cdot m+\frac{4d}{1-\gamma^2}+2
  \]
  bins.
  If $E_{\mathrm{grp}}$ contains no lower-level objects, including when $J=0$, we use no bins for this step.
  The denominator $5$ in the definition of $g_j$ allocates $\eta \cdot m/5$ to top-level objects and $4\eta \cdot m/5$ to lower-level objects, where the factor $4$ comes from the NFDH area bound.
  The resulting separate packing of $E_{\mathrm{grp}}$ uses at most
  \begin{equation}
  \label{eq:group-cost} \eta \cdot m+\delta^{\mathrm{width}}(d,\gamma)\text{ bins, where } \delta^{\mathrm{width}}(d,\gamma):=d+\frac{4d}{1-\gamma^2}+2.
  \end{equation}

  \paragraph{Running time.}
  The construction uses polynomially many one-object containment queries, followed by matching, chain grouping, and NFDH.
  The polynomial running-time exponent is independent of $\alpha,\beta,\gamma,\eta$.
  Dependence on these fixed parameters enters through the constants and segment sizes.
  The packing witness is used only in the analysis, while \eqref{eq:R} and \eqref{eq:group-cost} bound the number of containment groups and provide the explicit separate-packing guarantee required by Definition~\ref{def:roundable}.
\end{proof}

\subsection{Approximation factor calculations}
\label{app:parameter-budget}

We now substitute the parameter choices into the five charges.
The separator term satisfies
\[
  \frac{4}{\lambda\cdot(r-1)} \le \frac{\eps}{8}.
\]
The rounding contribution is $\nicefrac{\eps}{4}$.
Since $\lambda\le\nicefrac{1}{12}$ and $\eps<1$, the bounds on $\alpha$ from \eqref{eq:alpha-choice} give
\begin{equation}
\label{eq:alpha-derived}
\alpha\le\frac{1}{16}, \qquad
\text{and} \qquad
\alpha^3\le\frac{\lambda\cdot\eps}{256}.
\end{equation}
For own-grid cuts, \eqref{eq:alpha-choice} and \eqref{eq:alpha-derived} give $\alpha\le\nicefrac{\lambda\eps}{128}$ and $\alpha\le\nicefrac{1}{16}$, so
\[
  \frac{2}{1-\alpha} \cdot \frac{2\alpha}{\lambda} 
  = \frac{4\alpha}{\lambda\cdot(1-\alpha)} 
  \le \frac{4 (\nicefrac{\lambda\eps}{128})}{\lambda(\nicefrac{15}{16})}
  =\frac{\eps}{30} < \frac{\eps}{16}.
\]
For problematic children, $2/(1-\alpha)\le\nicefrac{32}{15}$ gives
\[
  \frac{2}{1-\alpha} \cdot \frac{\eps}{64} \le \frac{32}{15} \cdot \frac{\eps}{64} < \frac{\eps}{16}.
\]
Finally, \eqref{eq:alpha-derived} gives $\alpha^3\le\nicefrac{\lambda\eps}{256}$, and the same bound on $2/(1-\alpha)$ gives
\[
  \frac{2}{1-\alpha} \cdot \frac{4\alpha^3}{\lambda} 
  \le \frac{32}{15} \cdot \frac{4\eps}{256} 
  = \frac{\eps}{30} < \frac{\eps}{16}.
\]
Adding these three transfer terms to the separator and rounding charges, the total multiplicative approximation overhead is at most
\[
  \underbrace{\nicefrac{\eps}{8}}_{\text{separator}} +\underbrace{\nicefrac{\eps}{4}}_{\text{rounding}} +\underbrace{\nicefrac{\eps}{16}}_{\text{cuts}} +\underbrace{\nicefrac{\eps}{16}}_{\text{problematic}} +\underbrace{\nicefrac{\eps}{16}}_{\Zfringe} =\frac{9\eps}{16}.
\]
Consequently,
\[
  \frac{4}{\lambda(r-1)} 
  +\frac{\eps}{4}
  +\frac{2}{1-\alpha}\left(\frac{2\alpha}{\lambda}
  +\frac{4\alpha^3}{\lambda}
  +\frac{\eps}{64}\right)
  \le \frac{9\eps}{16} \le \eps.
\]
The separator contributes the additive constant $2$, representative rounding contributes $\delta_{\mathrm{grp}}$, and structured completion contributes $1/(1-\alpha)$.

\section{Concrete geometric classes and APTAS consequences}
\label{app:concrete}

This appendix proves the framework's concrete APTAS consequences for star-shaped and convex-polygon classes.
For a fixed finite family of star-shaped templates, we verify rasterizability by a boundary-neighborhood estimate and representative roundability by aligning star centers.
For convex classes, uniform fatness provides the grid condition, and bounded translation-containment width provides the rounding condition.
We then prove the converse relation between convex rasterizability and fatness, and construct two non-convex families showing why that equivalence does not extend to general polygonal regions.

\subsection{Fixed star-shaped templates}

The two interfaces use different template properties.
Their fixed boundary geometry bounds the area added by a grid-cell cover uniformly over positive scales and translations.
Star-shapedness makes aligned homothetic copies nested, so each template contributes one containment chain.

\begin{proof}[Proof of Corollary~\ref{cor:homothetic-star-ptas}]
  Let $Q_1,\ldots, Q_t$ be the fixed templates.
  For every $Q_a$, fix a rational point $c_a$ witnessing its star-shapedness.
  The nonempty kernel of a rational polygon contains such a point.
  Write $A_a:=\area(Q_a)$, let $b_a:=b(Q_a)$ be its canonical bounding-square side length, let $m_a$ be its number of boundary edges, and let $p_a$ be the sum of the $L_1$-lengths of those edges.
  These are positive rational constants determined by the fixed class.

  Consider $P=s(Q_a-c_a)+c_a+v$ for a positive rational scale $s$ and a translation $v$.
  Every point of $V_h(P)\setminus P$ lies within Euclidean distance at most $\sqrt{2} \cdot h\le2h$ of $\partial P$.
  Covering the $2h$-neighborhood of each boundary edge by its rectangular strip and two endpoint disks, and using $\pi\le4$, gives
  \[
    \area\bigl(V_h(P)\setminus P\bigr)
    \le4s\cdot p_a\cdot h+16m_a\cdot h^2.
  \]
  Define the rational class constant
  \[
    C:=\max\left\{1,
    \max_{1\le a\le t}
    \frac{4p_a\cdot b_a+16m_a\cdot b_a^2}{A_a}
    \right\}
    \qquad\text{and}\qquad
    \gridrate(\eta):=\nicefrac{\eta}{C}.
  \]
  Since $b(P)=s\cdot b_a$ and $\area(P)=s^2\cdot A_a$, the condition $h\le\gridrate(\eta)\cdot b(P)$, together with $\frac{h}{s\cdot b_a}\le1$, implies
  \begin{align*}
  \frac{\area(V_h(P)\setminus P)}{\area(P)}
   & \le \frac{4p_a\cdot b_a}{A_a}\frac{h}{s\cdot b_a}
  +\frac{16m_a\cdot b_a^2}{A_a}\left(\frac{h}{s\cdot b_a}\right)^2 \\
   & \le C\frac{h}{s\cdot b_a}
  \le\eta.
  \end{align*}
  Thus the class is uniformly rasterizable.

  If $0<\lambda\le\mu$, star-shapedness gives
  \[
    \lambda\cdot (Q_a-c_a)+c_a\subseteq\mu\cdot (Q_a-c_a)+c_a.
  \]
  Indeed, after dividing by $\mu$, every point on the left corresponds to a point on the segment from $c_a$ to a point of $Q_a$.
  Therefore, after aligning these witnessing centers, the homothetic copies of each template form one translation-containment chain.
  Thus the $t$ template chains give translation-containment width at most $t$.
  Lemma~\ref{lem:width-roundable} now supplies uniform representative roundability, and Theorem~\ref{thm:abstract-main} gives the claimed no-augmentation APTAS and rational packing.
\end{proof}

\subsection{Convex fatness and the grid condition}

For convex polygons, the area of an outer neighborhood is controlled by perimeter and radius.
A common fatness bound relates those quantities to the object's area, allowing one resolution to work for the whole class.

\begin{lemma}[Uniform fatness implies uniform rasterizability]
  \label{lem:fat-grid}
  Every uniformly $f$-fat class of convex polygons is uniformly rasterizable.
  For every rational $\bar f\ge f$, one valid grid-resolution function is $\Pi_{\bar f}(\eta):=\nicefrac{\eta}{16\bar f^2}$.
\end{lemma}

\begin{proof}[Proof of Lemma~\ref{lem:fat-grid}]
  Fix a shifted grid of cell side length $h$.
  Every point in a selected grid cell is at $L_\infty$-distance at most $h$, and hence at Euclidean distance at most $h\cdot \sqrt{2}$ from $P$.
  Therefore
  \[
    V_h(P)\subseteq P\mathbin{+}[-h,h]^2 \subseteq P\mathbin{+}\callB(0,\sqrt2\cdot h).
  \]
  For a planar convex body, Steiner's formula~\cite[\S4.2]{Sch14} bounds the area of that neighborhood, and the perimeter--diameter inequality bounds its linear term:
  \begin{align*}
  \area(V_h(P)\setminus P)
   & \le \sqrt2\cdot h\cdot \operatorname{per}(P)+2\pi\cdot h^2    \\
   & \le 2\pi\sqrt2\cdot r^{\mathrm{out}}(P)\cdot h+2\pi\cdot h^2.
  \end{align*}
  Since $b(P)\le D(P)\le2r^{\mathrm{out}}(P)\le2\bar f\cdot r^{\mathrm{in}}(P)$, the condition $h\le\Pi_{\bar f}(\eta)\cdot b(P)$ gives, after substituting the proposed grid-resolution function,
  \[
    \frac{h}{r^{\mathrm{in}}(P)} \le2\bar f\cdot \Pi_{\bar f}(\eta) =2\bar f\cdot \frac{\eta}{16\bar f^2} =\frac{\eta}{8\bar f}.
  \]
  Dividing by $\area(P)\ge\pi\cdot r^{\mathrm{in}}(P)^2$ yields
  \begin{align*}
  \frac{\area(V_h(P)\setminus P)}{\area(P)}
   & \le4\sqrt2\cdot \bar f^2\cdot \frac{\eta}{16\bar f^2}
  +8\bar f^2\cdot \left(\frac{\eta}{16\bar f^2}\right)^2           \\
   & =\frac{\sqrt2}{4}\cdot \eta+\frac{\eta^2}{32\bar f^2}
  \le\frac{\sqrt2}{4}\cdot \eta+\frac{\eta^2}{32}
  \le\eta.
  \end{align*}
  The estimate is uniform in the grid phase and depends only on the fixed rational bound $\bar f$, as Definition~\ref{def:gridregular} requires.
\end{proof}

Combining this grid estimate with the rounding construction from translation-containment width proves the convex-class corollary.

\begin{proof}[Proof of Corollary~\ref{cor:concrete}]
  Lemma~\ref{lem:fat-grid} supplies uniform rasterizability, and Lemma~\ref{lem:width-roundable} supplies uniform representative roundability under the corollary's width hypothesis.
  Theorem~\ref{thm:abstract-main} therefore gives the stated APTAS and its rational packing.
\end{proof}

The reverse implication uses the bulk inequality derived from rasterizability.
For a convex object, that area lower bound forces a lower bound on its minimum width, and hence on its inradius relative to its diameter.
This gives a uniform fatness bound.

\begin{proposition}[Uniform rasterizability and uniform fatness for convex classes]
  \label{prop:convex-grid-fat}
  For every class of rational convex polygons, uniform rasterizability is equivalent to uniform fatness.
  More precisely, if the class has a grid-resolution function $\gridrate$, then it is uniformly $\nicefrac{3}{\lambda}$-fat for the positive constant $\lambda$ in \eqref{eq:lambda-grid}.
\end{proposition}
\begin{proof}
  Uniform fatness implies uniform rasterizability by Lemma~\ref{lem:fat-grid}.
  Conversely, Proposition~\ref{prop:grid-bulk} gives $\area(P)\ge\lambda\cdot D(P)^2$ for every object $P$ in a uniformly rasterizable class.
  Let $\operatorname{width}(P)$ denote its minimum width, namely the minimum distance between two parallel supporting lines.
  A rectangle of side lengths $D(P)$ and $\operatorname{width}(P)$ contains $P$, so
  \[
    \area(P)\le D(P)\cdot \operatorname{width}(P), \qquad\text{and hence}\qquad \operatorname{width}(P)\ge\lambda\cdot D(P).
  \]
  The planar case of Steinhagen's inradius--width inequality~\cite{Ste22} gives $r^{\mathrm{in}}(P)\ge\nicefrac{\operatorname{width}(P)}{3}$, while $r^{\mathrm{out}}(P)\le D(P)$.
  Therefore
  \[
    \frac{r^{\mathrm{out}}(P)}{r^{\mathrm{in}}(P)} \le\frac{D(P)}{\operatorname{width}(P)/3} \le\nicefrac{3}{\lambda},
  \]
  which is a uniform fatness bound.
\end{proof}

For non-convex objects, a contained disk does not control the area added across narrow gaps, and a bound on outer-neighborhood growth does not guarantee a contained disk of comparable radius.
The first witness family has a fixed contained disk and increasingly narrow gaps that fill under an outer neighborhood.
The second has increasingly dense holes whose total boundary contribution remains bounded while every contained disk must shrink.

\begin{proposition}[Non-convex fatness and rasterizability are incomparable]
  \label{prop:nonconvex-grid-fat}
  There is a uniformly fat class of rational connected polygonal regions that is not uniformly rasterizable, and there is a uniformly rasterizable class of rational connected polygonal regions that is not uniformly fat.
\end{proposition}

\begin{figure}[t]
  \centering
  \begin{tikzpicture}[
      x=3.45cm,
      y=3.45cm,
      font=\small,
      line cap=round,
      line join=round,
      witness/.style={draw=aptasblue,fill=aptasbluelight!55,densely dashed,line width=0.65pt},
      object/.style={draw=aptasteal!80!black,fill=aptasteallight,line width=0.9pt}
    ]
    \path[use as bounding box] (-0.30,-0.22) rectangle (2.85,1.30);
    \begin{scope}
      \node[figtitle] at (0.50,1.2) {(a) comb $C_6$};

      \fill[aptasredlight!72] (0,0.5) rectangle (1,1);
      \fill[aptasteallight] (0,0) rectangle (1,0.5);
      \foreach \i in {0,...,5} {
        \pgfmathsetmacro{\toothleft}{\i/6}
        \pgfmathsetmacro{\toothright}{\i/6+1/36}
        \fill[aptasteallight] (\toothleft,0.5) rectangle (\toothright,1);
        \draw[draw=aptasteal!80!black,line width=0.9pt]
        (\toothleft,0.5)--(\toothleft,1)--(\toothright,1)--(\toothright,0.5);
      }
      \draw[draw=aptasteal!80!black,line width=0.9pt]
      (0,0.5)--(0,0)--(1,0)--(1,0.5);
      \foreach \i in {0,...,4} {
        \pgfmathsetmacro{\gapleft}{\i/6+1/36}
        \pgfmathsetmacro{\gapright}{(\i+1)/6}
        \draw[draw=aptasteal!80!black,line width=0.9pt]
        (\gapleft,0.5)--(\gapright,0.5);
      }
      \draw[draw=aptasteal!80!black,line width=0.9pt]
      ({5/6+1/36},0.5)--(1,0.5);
      \draw[draw=aptasgray,densely dashed,line width=0.55pt] (0,0) rectangle (1,1);

      \path[witness] (0.5,0.25) circle (0.25);
      \node[fignote] at (0.5,0.25) {$r=\nicefrac{1}{4}$};

      \draw[draw=aptasred!82!black,line width=0.45pt]
      ({1/36},1.01)--({1/36},1.085)
      ({1/6},1.01)--({1/6},1.085);
      \draw[<->,draw=aptasred!82!black,line width=0.65pt]
      ({1/36},1.055)--({1/6},1.055);
      \node[fignote,anchor=east,inner sep=1pt] at (0.4,1.055) {$<\nicefrac{1}{n}$};
    \end{scope}

    \begin{scope}[shift={(1.55,0)}]
      \node[figtitle] at (0.50,1.2) {(b) perforated square $H_5$};

      \fill[aptasteallight] (0,0) rectangle (1,1);
      \foreach \k in {1,...,4} {
        \pgfmathsetmacro{\gridcoord}{\k/5}
        \draw[figfine] (\gridcoord,0)--(\gridcoord,1);
        \draw[figfine] (0,\gridcoord)--(1,\gridcoord);
      }
      \foreach \i in {0,...,4} {
        \foreach \j in {0,...,4} {
          \pgfmathsetmacro{\holex}{(\i+0.5)/5}
          \pgfmathsetmacro{\holey}{(\j+0.5)/5}
          \path[fill=white,draw=aptasred!82!black,line width=0.55pt]
          ({\holex-0.022},{\holey-0.022}) rectangle ({\holex+0.022},{\holey+0.022});
        }
      }
      \draw[draw=aptasteal!80!black,line width=0.9pt] (0,0) rectangle (1,1);

      \draw[<->,draw=aptasred!82!black,line width=0.65pt] (0.1,0.1)--(0,0);
      \node[fignote,anchor=north,inner sep=1pt] at (0,-0.05) {$\frac{\sqrt2}{2n}$};

      \draw[figarrow] (0.80,-0.075)--(0.90,0.078);
      \node[fignote,anchor=north,inner sep=1pt] at (0.72,-0.1) {hole of side $a_n$};
    \end{scope}
  \end{tikzpicture}
  \caption{
    The two non-convex witness families in Proposition~\ref{prop:nonconvex-grid-fat}.
    Left: $C_n$ contains a disk of radius $\nicefrac{1}{4}$, while every upper gap is narrower than $\nicefrac{1}{n}$ and therefore lies in $(C_n)_\delta$ when $\delta\ge\nicefrac{1}{n}$.
    Right: $H_n$ has one removed square of side $a_n=\frac{1}{10n^3}$ at the center of every cell of an $n\times n$ grid.
    The holes contribute at most $4n^2\cdot a_n\cdot \delta$ to the outer-neighborhood bound, while every point of the unit square is within distance $\frac{\sqrt2}{2n}$ of a removed center.
    The panels show $n=6$ and $n=5$, respectively, and the hole widths in the right panel are enlarged for visibility.
  }
  \label{fig:nonconvex-incomparability}
\end{figure}
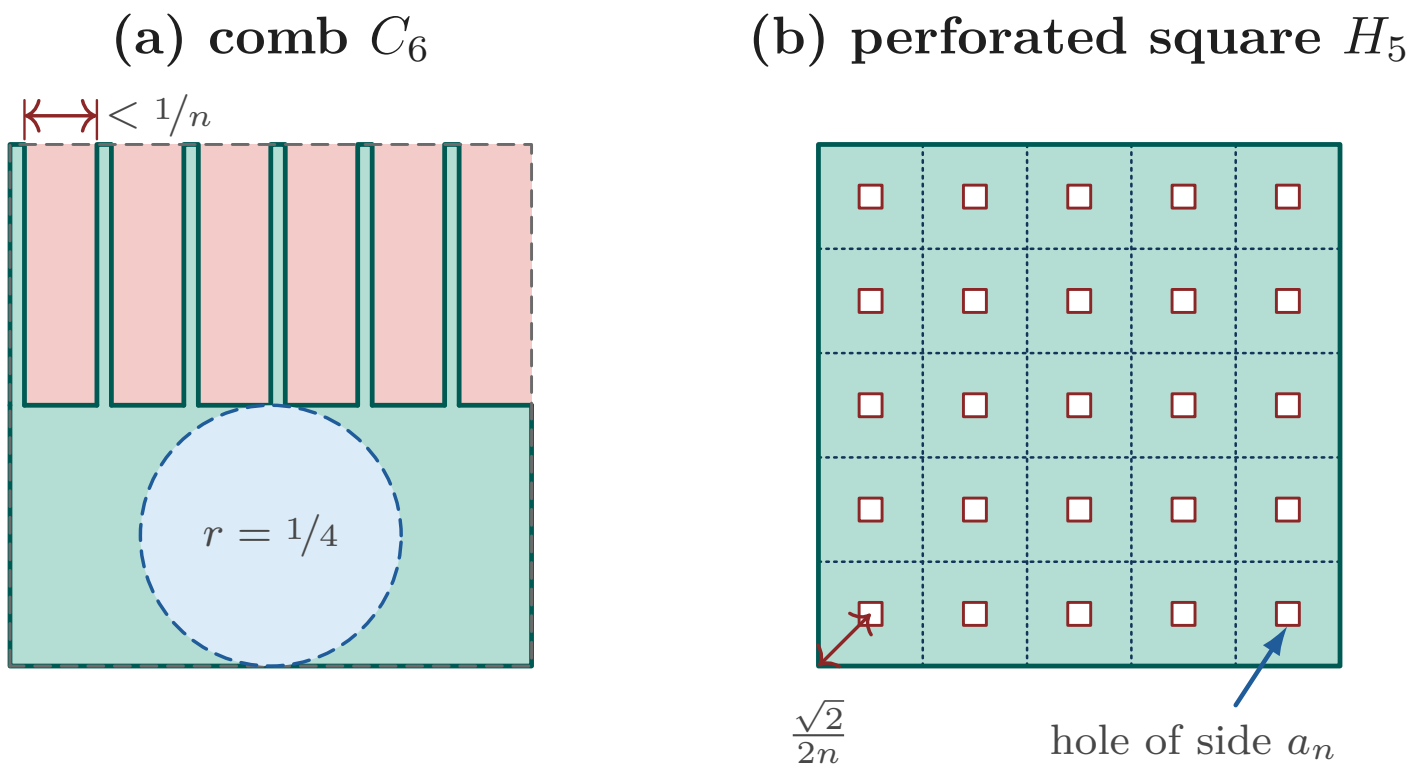

Figure~\ref{fig:nonconvex-incomparability} illustrates the two witness families used in the proof.

\begin{proof}
  For $n\ge4$, let
  \[
    C_n:=\left([0,1]\times\left[0,\frac{1}{2}\right]\right)
    \cup\bigcup_{i=0}^{n-1}
    \left(\left[\frac{i}{n},\frac{i}{n}+\frac{1}{n^2}\right]\times\left[\frac{1}{2},1\right]\right).
  \]
  This rational connected comb lies in the unit square and contains a disk of radius $\nicefrac{1}{4}$, so it is $3$-fat.
  Moreover, $b(C_n)=1$, $\area(C_n)=\nicefrac{1}{2}+\frac{1}{2n}$, and every upper gap has width less than $\nicefrac{1}{n}$.
  Hence, if $\delta\ge\nicefrac{1}{n}$, then $[0,1]^2\subseteq(C_n)_\delta$ and
  \[
    \area\bigl((C_n)_\delta\setminus C_n\bigr)
    \ge\frac{1}{2}-\frac{1}{2n}
    >\frac{1}{2}\area(C_n).
  \]
  A class-uniform outer-resolution value $\Sigma(\nicefrac{1}{2})>0$ would be contradicted by taking $\delta=\nicefrac{1}{2}\Sigma(\nicefrac{1}{2})$ and then $n$ with $\nicefrac{1}{n}\le\delta$.
  Proposition~\ref{prop:grid-parallel} therefore shows that $\{C_n:n\ge4\}$ is not uniformly rasterizable.

  For $n\ge1$, let $H_n$ be the unit square with the interiors of $n^2$ axis-aligned squares of side $a_n:=\frac{1}{10n^3}$ removed, one centered in each cell of the $n\times n$ grid.
  The holes lie strictly inside distinct cells, so $H_n$ is a rational connected polygonal region with $b(H_n)=1$ and $\area(H_n)=1-\frac{1}{100n^4}\ge\nicefrac{99}{100}$.
  For $0<\delta\le1$, the outer strip of the unit square and the inner strips of the holes give
  \[
    \area\bigl((H_n)_\delta\setminus H_n\bigr)
    \le4\delta+\pi\cdot \delta^2+4n^2\cdot a_n\cdot \delta
    \le8\delta.
  \]
  Thus $\Sigma(\eta)=\nicefrac{\eta}{16}$ is a uniform outer-resolution function, and Proposition~\ref{prop:grid-parallel} gives, for example, the rational grid-resolution function $\gridrate(\eta)=\nicefrac{\eta}{32}$.
  Finally, opposite corners give $r^{\mathrm{out}}(H_n)\ge\nicefrac{\sqrt2}{2}$, whereas every point of the unit square is within distance $\frac{\sqrt2}{2n}$ of a removed cell center, so $r^{\mathrm{out}}(H_n)/r^{\mathrm{in}}(H_n)\ge n$ and the class is not uniformly fat.
\end{proof}

\section{Structured packing details}
\label{app:structured-packing-details}

The geometric arguments below justify the comparison packing: completion restores all deleted copies, and deleting $\Zfringe$ leaves square regions.
We then compute exact capacities for local configurations and show how their stored placements compose into a packing.
Finally, counting occupied regions bounds the domains of configuration-count variables, allowing an integer program to optimize full-supply packings.
The last subsection establishes the program's encoding and running-time bounds.

\subsection{Completing the rounded supply}
\label{app:completion-details}

Completion has two packing stages.
First, scaled NFDH packs each lower level's rounded copies into square sub-bins of that level's side length.
Second, the integer scale ratios let us allocate those sub-bins from largest to smallest by subdividing unused squares, so their total area determines enough unit bins.
The geometric decay of the scales bounds the sum of the per-level NFDH overheads independently of the number of levels.
Here $Z$ is a submultiset of lower-level rounded copies, and its level-$j$ copies belong to $\widehat S_j$.
This analysis does not change the input levels $ S_j$.

\begin{proof}[Proof of Lemma~\ref{lem:structured-completion}]
  If $Z=\varnothing$, we use the empty packing with zero bins.
  In particular, $J=0$ requires no completion.
  Otherwise, for $1\le j\le J$, put $Z_j:=Z\cap\widehat S_j$ and $A_j:=\sum_{P\in Z_j}b(P)^2$.
  The scale window and \eqref{eq:bdiam} give $b(P)\le\alpha\cdot s_j$ for $P\in Z_j$.
  For nonempty $Z_j$, we apply Lemma~\ref{lem:nfdh} to its bounding squares after scaling all lengths by $1/s_j$.
  Scaling back gives $a_j$ nonempty square sub-bins of side $s_j$ with
  \begin{equation}
    \label{eq:completion-level-area}
    a_j\cdot s_j^2\le\frac{2A_j+s_j^2}{1-\alpha}.
  \end{equation}
  Indeed, the bound in that lemma with maximum relative side $\eta_j\le\alpha$ is $(2A_j/s_j^2+1)/(1-\eta_j)$.
  We take $a_j=0$ when $Z_j$ is empty and embed each labeled rounded copy inside its own bounding-square placement.

  Let $V:=\sum_{j=1}^J a_j\cdot s_j^2$ and $K:=\lceil V\rceil$.
  We start with $K$ unit bins and allocate the sub-bins from largest to smallest.
  Before allocating level $j$, we subdivide every unused level-$(j-1)$ square into $(s_{j-1}/s_j)^2$ level-$j$ squares.
  Squares holding an allocated sub-bin are never subdivided.
  Lemma~\ref{lem:preprocessing} implies the ratios, including $s_0/s_1$, are integers.
  Thus the number of unused level-$j$ squares is the integer
  \[
    \frac{K-\sum_{i<j}a_i\cdot s_i^2}{s_j^2}
    \ge\frac{\sum_{i\ge j}a_i\cdot s_i^2}{s_j^2}\ge a_j.
  \]
  We assign the $a_j$ sub-bins to distinct unused squares and continue, including through empty intermediate levels.

  We retain every ancestor of an allocated sub-bin and prune all unused branches and empty roots.
  Each allocated sub-bin becomes a level-$j$ region holding its level-$j$ rounded copies as current objects and having no children.
  Every other retained non-root region is an intermediate square with no current objects, and the roots also have no current objects.
  Subdivision gives contained, pairwise interior-disjoint children.
  Regions with current objects have no children, so children always avoid current objects.
  Every nonterminal region has at most $(s_j/s_{j+1})^2\le\Bchild$ children, and the bulk occupancy argument in \eqref{eq:K} bounds its current objects by $\Kloc$.
  Consequently, this is a structured packing of all labeled copies in $Z$ in at most $K$ bins.
  Grid alignment is used only for this analysis construction and does not restrict the movable child squares in local queries.

  The current scales satisfy $s_1\le\alpha^3$, $s_j=s_1\cdot \gamma^{j-1}$, $\gamma\le\nicefrac12$, and $\alpha\le\nicefrac1{16}$.
  Hence $\sum_{j=1}^J s_j^2\le\frac{s_1^2}{1-\gamma^2}\le\frac43\alpha^6\le\alpha$.
  Summing \eqref{eq:completion-level-area} and using $\lceil V\rceil\le V+1$ gives
  \[
    K\le\frac{2\sum_j A_j+\sum_j s_j^2}{1-\alpha}+1
    \le\frac{2}{1-\alpha}\sum_{P\in Z}b(P)^2+\frac{1}{1-\alpha}.
  \]
  Since $b(P)\le D(P)$, this implies the stated bound.
\end{proof}

\subsection{Common grid phase}
\label{app:global-shift}

\begin{proof}[Proof of Lemma~\ref{lem:globalshift}]
  For the analysis, we choose $\tau$ uniformly from $[0,s_1)^2$.
  For each level $j\ge1$, the ratio $s_1/s_j$ is an integer, so each coordinate of $\tau$ modulo $s_j$ is uniform in $[0,s_j)$.
  For a fixed level-$j$ object $P$, the measure of offsets placing a vertical grid line in the interior of its horizontal projection is at most the projection length, which is at most $D(P)$.
  The same bound holds for horizontal lines.
  A union bound therefore gives
  \[
    \Pr[P\text{ is cut}]\le\frac{2D(P)}{s_j}\le2\alpha.
  \]
  We sum area-weighted indicators over all lower-level rounded copies in the fixed packing.
  Their total area is at most $m$, so
  \[
    \mathbb E[\area(\Zcut)]\le2\alpha\cdot m.
  \]
  No independence between levels or objects is needed.
  Hence some common phase has $\area(\Zcut)\le2\alpha\cdot m$, and the bulk inequality $D(P)^2\le\area(P)/\lambda$ gives \eqref{eq:cut-bound}.
  This phase and its possibly irrational coordinates are used only to prove existence of a structured packing.
\end{proof}

\subsection{Deletion near bin boundaries}
\label{app:hierarchy-details}

For combinatorial ownership, the cell with indices $(z,z')$ is
\[
 [\tau_x+z\cdot h,\tau_x+(z+1)\cdot h)\times
 [\tau_y+z'\cdot h,\tau_y+(z'+1)\cdot h).
\]
This half-open convention gives unique indices, while geometric regions and covers use cell closures.
An uncut object's interior lies in one open cell, and the object lies in its closure because it is the closure of its interior.

\begin{figure}[H]
  \centering
  \begin{tikzpicture}[x=0.75cm,y=0.75cm,font=\small,line cap=round,line join=round]
    \begin{scope}
      \clip (1.65,1.65) rectangle (2.85,2.85);
      \foreach \x in {1.95,2.25,2.55} \draw[figfine] (\x,1.60)--(\x,2.90);
      \foreach \y in {1.95,2.25,2.55} \draw[figfine] (1.60,\y)--(2.90,\y);
    \end{scope}

    \foreach \x in {0.45,1.65,2.85,4.05} \draw[figgrid] (\x,0)--(\x,5.0);
    \foreach \y in {0.45,1.65,2.85,4.05} \draw[figgrid] (0,\y)--(5.0,\y);
    \draw[line width=1.05pt] (0,0) rectangle (5.0,5.0);

    \draw[activecell,fill opacity=0] (1.65,1.65) rectangle (2.85,2.85);
    \pic[currentpoly,xscale=1.8,yscale=1.45] at (2.25,2.31) {nonconvexPolygon};
    \node[font=\footnotesize] at (2.15,2.31) {$P$};
    \draw[decorate,decoration={brace,amplitude=3.5pt},draw=aptasblue] (1.65,3.00)--(2.85,3.00);
    \node[fignote] at (2.25,3.3) {$s_1$};

    \fill[discardcell] (0,0) rectangle (0.45,5.0);
    \fill[discardcell] (4.05,0) rectangle (5.0,5.0);
    \fill[discardcell] (0.45,0) rectangle (4.05,0.45);
    \fill[discardcell] (0.45,4.05) rectangle (4.05,5.0);

    \filldraw[draw=white,line width=0.7pt] (0.45,0.45) circle (0.075);
    \draw[figarrow,draw=black] (1.55,0.76)--(0.52,0.48);
    \node[fignote] at (1.9,0.86) {$\tau$};

    \draw[discardcell] (5.93,3.35) rectangle (6.48,3.90);
    \node[figlabel,anchor=west,align=left,text width=4.70cm] at (6.72,3.63)
    {boundary-clipped level-$1$ regions};
    \draw[activecell] (5.93,2.23) rectangle (6.48,2.78);
    \node[figlabel,anchor=west,align=left,text width=4.70cm] at (6.72,2.51)
    {a retained level-$1$ square};
    \draw[figfine] (5.93,1.25)--(6.48,1.25);
    \draw[figfine] (5.93,1.43)--(6.48,1.43);
    \draw[figfine] (6.11,1.08)--(6.11,1.61);
    \draw[figfine] (6.30,1.08)--(6.30,1.61);
    \draw[draw=aptasblue!72,line width=0.62pt] (5.93,1.08) rectangle (6.48,1.61);
    \node[figlabel,anchor=west,align=left,text width=5.45cm] at (6.72,1.35)
    {next-level grid tiling a retained level-$1$ square};
  \end{tikzpicture}
  \caption{
    Schematic of the level-$1$ grid and a finer grid for a fixed analysis phase $\tau$.
    After own-grid deletion, $\Zfringe$ consists of the remaining lower-level rounded copies whose level-$1$ ancestor regions are clipped by a bin boundary.
    Pruning unoccupied regions after deleting these copies leaves square regions at every retained depth.
  }
  \label{fig:globalphase}
\end{figure}
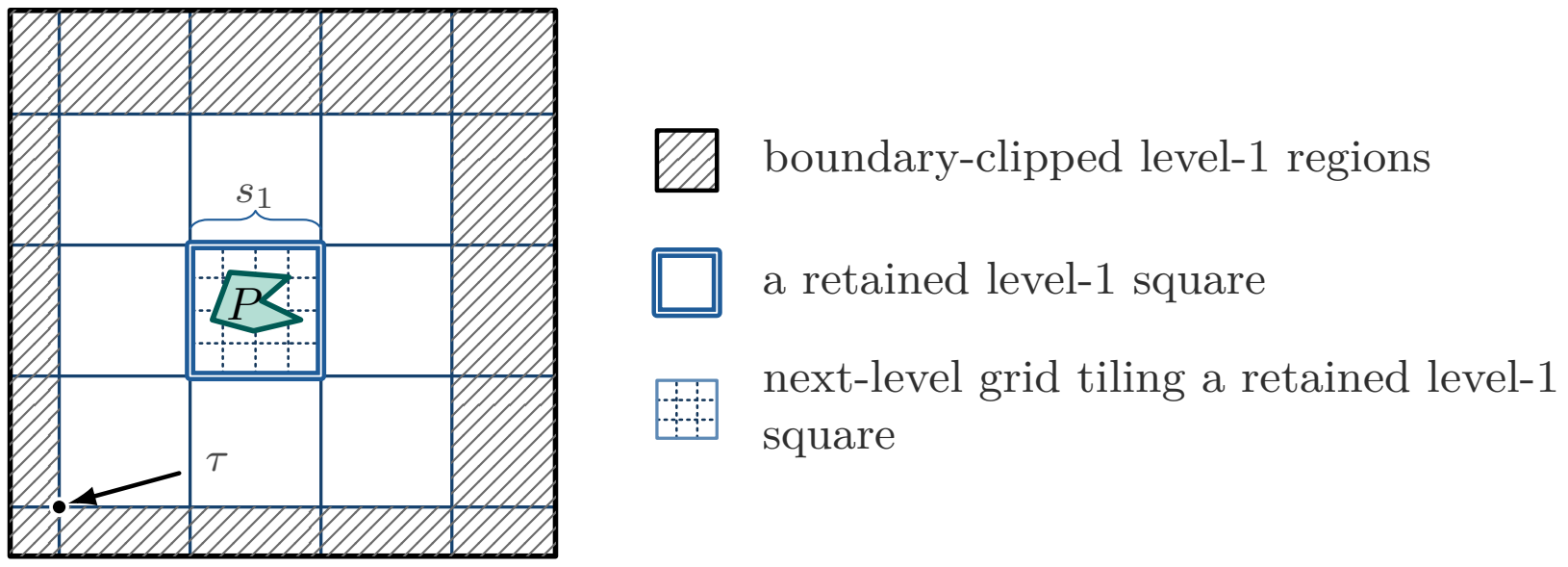

For a common phase $\tau\in[0,s_1)^2$, we use the grids $\mathcal G(s_j,\tau)$ at levels $j\ge1$ in each bin's unit-square coordinates.
Since $s_j/s_{j+1}=\Mgrid$ is an integer, every grid line is also a line of the next finer grid.
For a bin $B$, let $\operatorname{Gr}_{j,\tau}(B)$ consist of the intersections $B\cap\overline C$ for cells $C$ of $\mathcal G(s_j,\tau)$ satisfying $\operatorname{int}(C)\cap\operatorname{int}(B)\ne\varnothing$.
The bins and the regions in $\operatorname{Gr}_{j,\tau}(B)$ over all bins and lower levels form a rooted forest.
The bins are its roots, and half-open cell indices determine unique parents at consecutive levels.
Some regions are clipped by a bin boundary and need not be squares.
The analysis removes objects in those regions before obtaining a structured packing.
Figure~\ref{fig:globalphase} illustrates this auxiliary grid construction.

\begin{definition}
  \label{def:fringe}
  For a root region $B$ and phase $\tau$, let $F_\tau$ be the union of the clipped regions in $\operatorname{Gr}_{1,\tau}(B)$, namely those whose underlying $s_1\times s_1$ cell closures are not contained in $B$.
  After deleting $\Zcut$, the copies in $\Zfringe$ within $B$ are exactly the surviving rounded copies from levels $j\ge1$ whose level-$1$ ancestor region is contained in $F_\tau$.
\end{definition}

\begin{proof}[Proof of Lemma~\ref{lem:square-hierarchy}]
  In one root, every clipped level-$1$ region lies in the union of four boundary strips of width less than $s_1$, whose total area is at most $4s_1$.
  Every copy in $\Zfringe$ lies in one of these clipped regions and hence in the union of the four strips.
  These rounded copies belong to one packing and have pairwise disjoint interiors.
  Their total area over all roots is at most $4s_1m$.
  Uniform bulk gives
  \[
    \sum_{P\in\Zfringe}D(P)^2 \le\frac{\area(\Zfringe)}{\lambda} \le\frac{4s_1}{\lambda}m.
  \]
  Also $s_1=\alpha^{2t+1}\le \alpha^3$ as $t\ge1$, which proves the second inequality in \eqref{eq:fringe-bound}.
\end{proof}

For the structural claims in Section~\ref{sec:transfer}, note that every retained level-$1$ region is a full square.
For $1\le j<J$, each retained level-$j$ square is tiled by exactly $\Mgrid^2$ cell closures of the next grid.
Induction gives square descendants and at most $\Mgrid^2$ occupied children per non-root region.
A root contains at most $s_1^{-2}=\alpha^{-2(2t+1)}<\alpha^{-4r}=\Mgrid^2$ level-$1$ cell closures because $t\le r-1$.

\subsection{Problematic children}
\label{app:problematic-children}

\begin{proof}[Proof of Lemma~\ref{lem:problematic}]
  For every current object $P\in\widehat S_j$ at a nonterminal level, Lemma~\ref{lem:preprocessing} gives $h_j:=s_{j+1}\le\gridrate(\nicefrac{\lambda\eps}{64})\cdot b(P)$.
  Uniform rasterizability therefore bounds $\area(V_{h_j}(P)\setminus P)$ by $\nicefrac{\lambda\eps}{64}\area(P)$.

  Let $\mathcal C_P$ be the children charged to $P$.
  The deleted descendants below $C\in\mathcal C_P$ lie in $C$ and are interior-disjoint from $P$, so their area is at most $\area(C\setminus P)$.
  Each charged child square is the closure of a child-grid cell whose interior meets $\operatorname{int}(P)$, so it is contained in $V_{h_j}(P)$.
  Distinct charged child squares have disjoint interiors.
  Thus their total charge is at most $\nicefrac{\lambda\eps}{64}\area(P)$, including when $P$ has holes.

  The selected children form an anti-chain because processing stops below each deletion, so every removed object is charged once.
  The charged current objects are distinct members of the fixed packing and have total area at most $m$, so $\area(\Zprob)\le\nicefrac{\lambda\eps}{64}\cdot m$.
  Bulk gives \eqref{eq:problem-bound}.
\end{proof}

\subsection{Computing exact local capacities}
\label{app:capacity-details}

The occupancy bound limits the total current multiplicity, while rounding limits the number of coordinates in a configuration.
These bounds give the number of local queries below.
For each feasible current configuration, deleting child squares preserves feasibility, so we can search for its maximum child count and retain a realization attaining it.

Padding the multiplicity vector to $\Rrep$ coordinates and adding a slack coordinate gives the stars-and-bars bound $|\mathcal X_j|\le\Hpat=\binom{\Rrep+\Kloc}{\Kloc}$.
Thus $\Hpat$ depends only on $\calP$ and $\eps$.

Using Lemma~\ref{lem:local-solver}, we can compute the feasible configurations and their capacities, together with one maximizing rational packing realization per configuration, at each level in time
\[
  \Hpat\,O\bigl(1+\log(\Bchild+1)\bigr)\,F_{\mathrm{loc}}(\Kloc+\Bchild)L_{\mathrm{loc}}^{O(\Kloc+\Bchild)}.
\]
Here $L_{\mathrm{loc}}$ is the maximum binary encoding length among the generated local instances at that level.
Since each query contains at most $\Kloc+\Bchild$ objects, bounded only by $\calP$ and $\eps$, this gives the accuracy-dependent input-size exponent allowed for an APTAS.

\begin{lemma}
  \label{lem:exact-capacity}
  For fixed accuracy, we can compute $\mathcal X_j$, its capacities, and one rational maximizing realization per configuration in time polynomial in $L$.
\end{lemma}

\begin{proof}
  We enumerate all nonnegative integer multiplicity vectors $\mathbf{c}$ with $|\mathbf{c}|_1\le\Kloc$.
  For each vector, we first invoke Lemma~\ref{lem:local-solver} on its rounded copies alone in $R_j$.
  If this query is infeasible, we discard $\mathbf{c}$: adding child squares cannot restore feasibility.
  Otherwise $\mathbf{c}\in\mathcal X_j$, so we store the returned packing realization for child count zero.
  At level $J$, this completes the computation and the capacity is zero.

  For $j<J$, we query a count $q$ by adding $q$ identical axis-aligned squares of side $s_{j+1}$ to the rounded copies.
  Feasibility is downward closed in $q$, since deleting child squares preserves a packing, and area gives $\kappa_j(\mathbf{c})\le\Bchild$.
  Binary search over $0\le q\le\Bchild$ finds $\kappa_j(\mathbf{c})$.
  We retain a maximizing realization with indexed child-square placements.
  For the empty configuration, tiling $R_j$ by child squares attains the area bound $\left(\nicefrac{s_j}{s_{j+1}}\right)^2$.

  The initial call and the binary search use $O(1+\log(\Bchild+1))$ calls per configuration.
  Every call has at most $\Kloc+\Bchild$ objects, including the auxiliary squares, and at most $\Hpat$ configurations are considered.
  Lemma~\ref{lem:local-solver} therefore gives the claimed running time and polynomial-bit rational packing realizations.
\end{proof}

\subsection{Geometric composition}
\label{app:composition-details}

\begin{proof}[Proof of Lemma~\ref{lem:composition}]
  In a structured packing, every occupied child square is interior-disjoint from the current objects in its parent.
  Together with the current rounded copies, these squares form a feasible query for their child count, so the definition of $\kappa_j(\mathbf{c})$ proves the first assertion.

  For the converse, we process an abstract forest from leaves to roots, using stored realizations to place current copies and assign assembled child packings injectively to reserved child squares.
  We translate each child packing into its assigned square, omitting unused child squares.
  Every child sub-bin at level $j+1$ is a translate of $[0,s_{j+1}]^2$, so translating its entire assembled packing preserves every descendant containment and disjointness relation.
  The reserved child squares and current rounded copies have pairwise disjoint interiors, proving feasibility by induction.
  We use a unit bin at each root, then prune unoccupied nodes, including empty roots.
  This preserves group multiplicities and gives the asserted bin bound.
  Reconstructing configurations extracted from a packing need not preserve its coordinates and does not give a bijection between packings and configuration choices.
\end{proof}

\subsection{Counting occupied regions}
\label{app:occupied-details}

The global optimization counts regions of each configuration, so it needs an upper bound on how many regions can occur at a level.
Empty branches contribute no objects and can be removed.
Every remaining region can be associated with an object in its subtree.
This gives domains bounded by the input cardinality for the integer variables introduced next.

Given any structured packing, we remove every region whose subtree contains no assigned object, including an entire bin if it is empty.
This operation does not move an object or change a multiplicity for any containment group, and every remaining child still occupies one of its parent's distinct reserved squares.
An empty current configuration remains when its region routes an assigned object from a deeper level.
Let $N_j^{\mathrm{occ}}$ be the number of remaining level-$j$ regions.
For each remaining region at a fixed level $j$, choose one assigned object in its subtree.
Distinct level-$j$ regions have disjoint subtrees, so these chosen objects are distinct.
Therefore
\begin{equation}
  \label{eq:occupied-bound}
  N_j^{\mathrm{occ}}\le n\qquad(0\le j\le J).
\end{equation}
There are at most $n(J+1)$ remaining regions over all levels.
For empty supply, the empty forest has zero bins.

\subsection{Integer program for full-supply structured packing}
\label{app:selection-proof}

Geometric composition permits us to record how often each local configuration is used, leaving the stored realizations to supply coordinates during reconstruction.
The program enforces exact demands for every containment group and enough aggregate capacity between each pair of consecutive levels.
Because the child squares at a level have the same size, each assembled child packing can use any reserved square of that level.
The equivalence proof below turns the count inequalities into an abstract forest and then realizes a structured packing containing only occupied regions.
The demands $n_{j,u}=|\Sclass_{j,u}|$ count input objects in a containment group and, equivalently, labeled copies of its chosen maximum in $\widehat S_j$.
All of these copies must be packed.

\begin{proof}[Proof of Lemma~\ref{lem:structured-selection}]
  For every level $j$ and configuration $\mathbf{c}\in\mathcal X_j$, let the integer variable $x_{j,\mathbf{c}}$ count the regions using that configuration.
  Write the region count, child capacity, and multiplicities for the containment groups as the linear expressions
  \[
    N_j(x):=\sum_{\mathbf{c}\in\mathcal X_j}x_{j,\mathbf{c}},
    \qquad C_j(x):=\sum_{\mathbf{c}\in\mathcal X_j}\kappa_j(\mathbf{c})\cdot x_{j,\mathbf{c}},
    \qquad q_{j,u}(x):=\sum_{\mathbf{c}\in\mathcal X_j}c_u\cdot x_{j,\mathbf{c}}.
  \]
  We impose exact demands for every containment group, adjacent-level capacities, and the occupied-region bounds:
  \begin{align}
    q_{j,u}(x)&=n_{j,u} &&(\text{every containment group }\Sclass_{j,u}), \notag\\
    N_{j+1}(x)&\le C_j(x) &&(0\le j<J), \label{eq:selection-ip-constraints}\\
    N_j(x)&\le n &&(0\le j\le J), \notag\\
    0\le x_{j,\mathbf{c}}&\le n,\qquad x_{j,\mathbf{c}}\in\mathbb Z &&(0\le j\le J,\ \mathbf{c}\in\mathcal X_j). \notag
  \end{align}
  These expressions introduce no additional variables.
  We solve the integer linear program $\min\bigl\{N_0(x):\eqref{eq:selection-ip-constraints}\bigr\}$ using the bounded-domain optimization algorithm justified below.

  A structured packing of $\widehat S$ always exists in at most $|\widehat S|\le n$ bins. For each level-$j$ copy, we use a separate bin with a chain of lower-left-aligned regions of side lengths $s_0,\ldots,s_j$, assigning the copy only to the last region.
  Top-level copies fit a unit bin by assumption, and lower-level copies fit their regions since $b(P)\le D(P)\le\alpha\cdot s_j<s_j$.

  Given any structured packing covering the rounded supply, we prune regions whose subtrees contain no assigned object.
  Its configuration counts satisfy \eqref{eq:selection-ip-constraints}: every containment group's demand is met exactly, Lemma~\ref{lem:composition} gives the child-capacity inequalities, and \eqref{eq:occupied-bound} gives the region bounds.
  Its number of bins equals $N_0(x)$, so the integer optimum is no greater than the minimum number of bins in a full-supply structured packing.

  Conversely, given a feasible integer solution, we create $x_{j,\mathbf{c}}$ abstract nodes of configuration $\mathbf{c}$ at each level.
  For each adjacent pair of levels, the capacity inequality permits an injective assignment of all next-level nodes to reserved child squares in the preceding nodes' stored realizations, forming an abstract rooted forest.
  The demand equalities give exactly $n_{j,u}$ current rounded-copy placements for each containment group $\Sclass_{j,u}$.
  We assign the copies bijectively within each containment group.
  Lemma~\ref{lem:composition} assembles the stored rational packing realizations and prunes empty branches, preserving every copy and giving a full-supply structured packing in at most $N_0(x)$ bins.
  The two directions prove equality of the integer optimum and the minimum structured bin count.
  At an optimum, pruning cannot strictly reduce that count.

  We now choose, within each containment group, any bijection between its input objects and the corresponding rounded copies.
  For each input object, we compute a rational containment translation into its chosen maximum by a single-object query to Lemma~\ref{lem:local-solver}.
  We add this translation to the translation of its assigned rounded copy.
  Containment preserves all disjointness relations, and the level-separated partition and distinct input labels ensure that every input object in the containment groups appears exactly once.
  The returned packing of the input objects in the containment groups uses the same bins.
  The optimality assertion concerns the rounded structured problem, since substitution may permit other packings of those input objects.

  The empty configuration is available even at a level with no supply, allowing regions to route deeper objects through that level.
  If the rounded supply is empty, the zero vector is feasible and optimal, yielding the empty packing.
  If $J=0$, there is one configuration block with exact demands for every containment group and no transition constraints.
\end{proof}

\subsection{Encoding and running time}
\label{app:selection-runtime}

The number of levels can grow with the input, but each level contributes only a parameter-bounded block of variables.
Constraints connect variables within a block or across two consecutive blocks, which gives the path decomposition used by the bounded-domain algorithm.
We first establish the width bound and then account for the program encoding, coordinate reconstruction, and local-query running time.

The primal graph of an integer program has one vertex per variable and joins two variables whenever they occur together in a constraint.
Let $B_j:=\{x_{j,\mathbf{c}}:\mathbf{c}\in\mathcal X_j\}$ and $k:=\max_j|B_j|\le\Hpat$.
The empty configuration ensures $k\ge1$.
All constraints in \eqref{eq:selection-ip-constraints} lie in one block $B_j$ or in two consecutive blocks, so for $J\ge1$ the bags $W_j:=B_j\cup B_{j+1}$, $0\le j<J$, form a path decomposition for the configuration variables.
For $J=0$, we use the single bag $B_0$.
The bags containing any variable are consecutive, and each bag has size at most $2k$ (at most $k$ when $J=0$).
Thus the program has primal pathwidth at most $2k-1$.

We convert this path decomposition into a nice Gaifman decomposition in polynomial time without increasing its width, using Jansen and Kratsch's construction~\cite[Proposition~2 in the full version]{JK15}.
Their bounded-domain algorithm, given a nice Gaifman decomposition of width at most $w:=2k-1\ge1$ and domains of size at most $d$, decides integer feasibility in $O(d^{w+1}w(N_{\mathrm{IP}}+M_{\mathrm{IP}}))$ arithmetic operations, where $N_{\mathrm{IP}}$ and $M_{\mathrm{IP}}$ are the numbers of variables and constraints~\cite[Theorem~3 in the full version]{JK15}.
For linear optimization, we append forget nodes to empty the root bag without increasing the width, and we store minimum objective contributions in feasible states.
We charge each variable's cost when it is forgotten.
Retaining minimizing choices recovers an optimizer.
This uses the same states and a polynomial bit-arithmetic overhead for integer costs.
The objective is handled by these costs and introduces no additional constraint joining variables across levels.
Every variable in our program has at most $n+1$ possible values, so one optimization call takes
\begin{equation}
  \label{eq:selection-ip-runtime}
  (n+1)^{2k}\cdot L_{\mathrm{IP}}^{O(1)},
\end{equation}
where $L_{\mathrm{IP}}$ is the program's binary encoding length, including its integer objective.

The scale construction and polynomial encoding bounds are proved in Appendix~\ref{app:scale-construction}.
Only the parameter-bounded level-$0$ block has nonzero objective coefficients, all equal to one, and its variables are bounded by $n$, so accumulated objective values have $O(\log(n+1))$ bits for fixed accuracy.
The program has at most $(J+1)\cdot \Hpat$ variables and $O((J+1)\cdot (\Hpat+\Rrep))$ constraints, each with parameter-bounded support, so $L_{\mathrm{IP}}\le F(\nicefrac{1}{\eps})\cdot L^{O(1)}$.
Each local instance has encoding length at most $F(\nicefrac{1}{\eps})\cdot L^{c_0}$ for an absolute $c_0$, and the local solver returns polynomial-bit rational translations.
Reconstruction uses at most one single-object containment query per input object in a containment group.
These queries produce polynomial-bit rational containment translations and take polynomial time with an exponent independent of the accuracy parameters.
A region path adds at most $J+1$ child-square offsets to one local and one containment translation, preserving a polynomial encoding bound with a uniform exponent.
At most $n$ occupied regions per level are instantiated.
Combining \eqref{eq:selection-ip-runtime} with Lemma~\ref{lem:exact-capacity}, the polynomial-time scale construction and separator packing, the polynomial-time representative-rounding interface, the containment queries for reconstruction, and the parameter-bounded number of separator residues give $L^{g_{\calP}(\nicefrac{1}{\eps})}$ for a computable function $g_{\calP}$.

\clearpage
\bibliographystyle{alpha-compact}
\bibliography{bib}

\end{document}